\documentclass[aps,floats,prb,showpacs,twocolumn]{revtex4}

\usepackage{amsfonts,amsmath,mathrsfs} \usepackage{bm} \usepackage{bbm} \usepackage{dcolumn}
\usepackage{epsfig} \usepackage{latexsym}

\begin{document}

\title{Controlling transmission eigenchannels in random media by edge reflection}

\author{Liyi Zhao$^1$, Chushun Tian$^1$, Yury P. Bliokh$^2$, and Valentin Freilikher$^3$}

\affiliation{$^1$Institute for Advanced Study, Tsinghua University, Beijing, 100084, China\\
$^2$Department of Physics, Technion-Israel Institute of Technology, Haifa 32000, Israel\\
$^3$Department of Physics, Bar-Ilan University, Ramat-Gan 52900, Israel}


\date{\today}

\begin{abstract}

Transmission eigenchannels and associated eigenvalues, that give a
full account of wave propagation in random media,
have recently emerged as a major theme in theoretical and applied optics. Here we demonstrate, both
analytically and numerically, that in quasi one-dimensional ($1$D) diffusive samples,
their behavior is governed mostly by the asymmetry in
the reflections of the sample edges rather than by the absolute values of
the reflection coefficients themselves. We show that there exists a
threshold value of the asymmetry parameter, below which high transmission eigenchannels exist, giving rise to a singularity in the distribution of the transmission eigenvalues, $\rho({\cal T}\rightarrow 1)\sim(1-{\cal T})^{-\frac{1}{2}}$. At the threshold, $\rho({\cal T})$ exhibits critical statistics with a distinct singularity $\sim(1-{\cal T})^{-\frac{1}{3}}$; above it the high transmission eigenchannels disappear and $\rho({\cal T})$ vanishes for ${\cal T}$ exceeding a maximal transmission eigenvalue.
We show that such statistical behavior of the transmission
eigenvalues can be explained in terms of effective
cavities (resonators), analogous to those in which the states are trapped in
$1$D strong Anderson localization. In particular, the $\rho ( \mathcal{T}) $-transition can be mapped onto the shuffling of the resonator with perfect transmittance from the sample center to the edge with stronger reflection. We also find a similar transition in the distribution of resonant transmittances in $1$D layered samples.
These results reveal a physical connection
between high transmission eigenchannels in diffusive systems and $1$D strong Anderson localization. They open up a fresh
opportunity for practically useful application: controlling the transparency
of opaque media by tuning their coupling to the environment.

\end{abstract}


\maketitle

\section{Introduction}
\label{sec:introduction}

Wave progresses in random media exhibits rich physics \cite{Akkermans07,Sheng90,Lagendijk09}, which finds numerous practical applications ranging from electron devices
to optical communications and imaging.
When a wave is incident on an open random medium it is decomposed into a number of ``partial waves'' that propagate independently along natural channels -- the so-called transmission eigenchannels -- and are superposed again when they leave the medium. These channels can be obtained from the transmission matrix ${\bf t}$, whose elements are coefficients of field transmission through random media. The singular value decomposition of this matrix, ${\bf t}=\sum_n {\bf u}_n\sqrt{\tau_n}{\bf v}_n$, gives the waveforms at the input and output edges of the $n$th transmission eigenchannel, i.e., the unit vectors ${\bf u}_n$ and ${\bf v}_n$, respectively and the
corresponding transmission eigenvalues $\tau_n$ \cite{Genack15}. The transmission eigenvalues and the eigenchannel structure -- the corresponding spatial profiles of the wave fields inside the medium -- give a full account of wave propagation in the interior of open random media. In recent years, the coherent control of the incident classical wave field has made it possible to control transmitted waves (e.g., Refs.~\onlinecite{Mosk08,Lagendijk12,Derode95,Lagendijk10,Popoff10,Davy12,Genack12,Kim12,Cao14,Cui14,Aubry14,Cao15}), which has substantially advanced studies of wave propagation in random media. Subsequent investigations have been extended from the traditional subject of global transport behavior (e.g., transmission from the input to output edge) to the new realm of the eigenchannel structure (see Ref.~\onlinecite{Genack15a} for a review).

Edge reflection arises from the refractive index mismatch at the
surface of a sample, and is ubiquitous in all dielectric materials. The importance of edge reflection to the study diffusive wave transport was first pointed out in Ref.~\onlinecite{Lagendijk89}. However, the
investigations of its impact have so far largely been restricted to the average transmission and intensity of waves (see, e.g., Refs.~\onlinecite{Lagendijk89,Zhu91,Luck93,Genack93} as well as Ref.~\onlinecite{Rossum99} for a review). On the other hand, by reinjecting radiation that arrives at the edge from the interior of the sample, edge reflection strongly affects the coherence of waves and influences
dramatically the transmission eigenchannels and eigenvalues at each
individual realization. This fundamental issue yet remains largely unexplored.

\begin{table*}
\newcommand{\tabincell}[2]
{\begin{tabular}{@{}#1@{}}#2
\end{tabular}}
\centering
\caption{\label{Table3} Main characteristics of various phases.}
\begin{tabular}{c
cc}
  \hline\hline
   phase &
   maximum transmission eigenvalue\,\,\,\, & asymptotic behavior of $\rho({\cal T})$ \\
  \hline
  $O$ \,\,&
  ${\cal T}_{\rm max}=1$ & $\rho({\cal T}\rightarrow 1)\sim (1-{\cal T})^{-\frac{1}{2}}$\\
  \hline
  $C$ \,\,&
  ${\cal T}_{\rm max}<1$ & $\rho({\cal T}\geq {\cal T}_{\rm max})=0$\\
  \hline
  Critical \,\,&
  ${\cal T}_{\rm max}=1$ & $\rho({\cal T}\rightarrow 1)\sim (1-{\cal T})^{-\frac{1}{3}}$\\
  \hline\hline
\end{tabular}
\end{table*}

In a recent study \cite{Tian13} of wave transport through quasi
one-dimensional ($1$D) diffusive media, it was discovered that an interesting ``phase
transition'' occurs as the reflection of
one sample edge increases while the other edge remains
transparent. This phase transition is seen in the
distribution of transmission eigenvalues (DTE), defined as $\rho({\cal T})=\langle \sum_n \delta ({\cal T}-\tau_n)\rangle$
with $\langle\cdot\rangle$ being the disorder average. In the absence of edge reflection,
\begin{equation}\label{eq:20}
    \rho({\cal T})=
    \frac{\xi}{2L} \frac{1}{{\cal T}\sqrt{1-{\cal T}}}\equiv \rho_0({\cal T}),
\end{equation}
where $L$ is the sample length \cite{note1} and $\xi (\gg L)$ the localization length. This bimodal distribution was obtained in eighties of the past
century, by Dorokhov \cite{
Dorokhov82} and Mello, Pereyra, and Kumar
\cite{Mello88} (see Ref.~\onlinecite{Beenakker97} for a review on the early status of this distribution)
and has recently received considerable renewed interest \cite{Stone13,Genack12,Aubry14,Tian13,Cao15}. It lays a foundation for mesocopic physics of electrons and photons. The singularity of $\rho_0({\cal T})$ for ${\cal T}\rightarrow 1$ reflects the presence of high transmission eigenchannels that dominate wave transport \cite{Imry86}. An important question is: would these eigenchannels be blocked by edge reflection? In Ref.~\onlinecite{Tian13}, it was found that if edge reflection is below a certain critical value, they still exist, and the DTE displays the same singularity as $\rho_0({\cal T})$. Above the critical value, perfectly transmitting eigenchannels disappear and the DTE is unimodal. At the critical value, the system undergoes a sharp transition and critical statistics emerges, displaying a distinct singularity. This exact result completely washes out the common belief, i.e., that the main effect of edge reflection is to elongate the sample length as enforced by the well-known result of the average total transmission \cite{Lagendijk89,Zhu91,Genack93} obtained by using
either the diffusion model or radiative transfer theory \cite{Rossum99,Morse53,Chandrasekhar}. Instead, it suggests that even for diffusive waves, edge reflection affects
significantly properties of transmission eigenchannels and eigenvalues.

Many fundamental questions are thereby opened up. 
(i) How universal is the DTE transition? In fact,
the finding of Ref.~\onlinecite{Tian13} relies on an exact solution, 
obtained owing largely to the system's simple (and somewhat artificial) construction. 
That is, one sample edge is transparent and the other reflective. 
A question naturally arises: what happens to realistic systems where both sample edges are generally semitransparent? 
(Studies of the complex eigenvalues of the non-Hermitian Hamiltonian corresponding to these systems have shown that asymmetry 
of edge coupling to the environment leads to interesting transport phenomena \cite{Zelevinsky10}.) For such systems,
would the phase transition still occur and the universality (e.g., the critical statistics of transmission eigenvalues) be affected? (ii) The origin of the transition has so far remained unclear. Notably, whether and how does wave interference gives rise to this transition?

The purpose of this work is to extensively explore these subjects. To this end we consider wave transport through quasi $1$D diffusive samples with {\it arbitrary} edge reflectivities. We show analytically and numerically that the asymmetry of the surface
interaction at two edges of the sample leads to rich phase transition phenomena. Specifically, for the asymmetry parameter below a certain threshold value -- even though
the reflections of both edges could be strong -- perfectly
transmitting eigenchannels are opened,
and the DTE exhibits a $(1-{\cal T})^{-\frac{1}{2}}$
singularity (dubbed ``{\it O}-phase'').
Above the threshold, these channels are closed
(dubbed ``{\it C}-phase'') and the DTE becomes unimodal,
vanishing above the maximum eigenvalue ${\cal T}_{\rm max}$ ($<1$).
At the threshold, critical statistics with a $(1-{\cal T})^{-\frac{1}{3}}$ singularity occurs (dubbed ``Critical phase'').
The main properties of these three phases are summarized in Table~\ref{Table3}.
As we will show below, these asymptotic behaviors are insensitive to the system's details and thereby universal.
Our analytical and numerical analysis suggest that the perfectly transmitting (with the eigenvalue ${\cal T}=1$) eigenchannel is associated with an effective
resonator bound to the center of the corresponding eigenchannel profile. Quite surprisingly, these resonators, although exist in diffusive samples,
are of the same physical nature as the effective cavities, in which the
eigenstates of the strongly disordered $1$D systems are localized
\cite{Freilikher03}. Much as in the case of strong localization, the effective resonators
in diffusive samples are formed by disorder-induced barriers and are of
a small size -- of order of the transport mean free path. When the
asymmetry parameter increases, the centers of the eigenchannel profile
and the resonator move towards the sample edge with stronger reflection. As
they arrive at the edge, the perfectly transmitting eigenchannel
disappears and the DTE transition occurs.

We remark that the bimodal distribution (\ref{eq:20}) has been derived by various theoretical methods in the past three decades \cite{Dorokhov82,Mello88,Nazarov94,Frahm95,Rejaei96,Zirnbauer04,Tian05} and very recently received experimental confirmation \cite{Aubry14}.
In this work, we show that, physically, the square root
singularity in this distribution can be attributed to that the resonators are homogeneously distributed inside the
sample.

The remainder of this paper is organized as follows. In Sec.~\ref{sec:summary}, we give a brief digest of the main results, and qualitatively discuss
their physical meaning. In Sec.~\ref{sec:transition_diffusive}, we present an analytical theory for DTE of quasi $1$D disordered systems. In Sec.~\ref{sec:numerical result}, a numerical
verification of the analytical results is given. In Sec.~\ref{sec:physical_interpretation}, we introduce the
$1$D resonators model, and use it to explain the physics of statistical
behavior of the transmission eigenvalues in diffusive samples of higher
dimension. We conclude and briefly discuss the results in Sec. in Sec.~\ref{sec:conclusions}. Some technical details are given in Appendices \ref{sec:calculations}-\ref{sec:parametrization}.

\section{Summary of main results and their physical meanings}
\label{sec:summary}

\begin{figure}[h]
  \centering
\includegraphics[width=8.0cm]{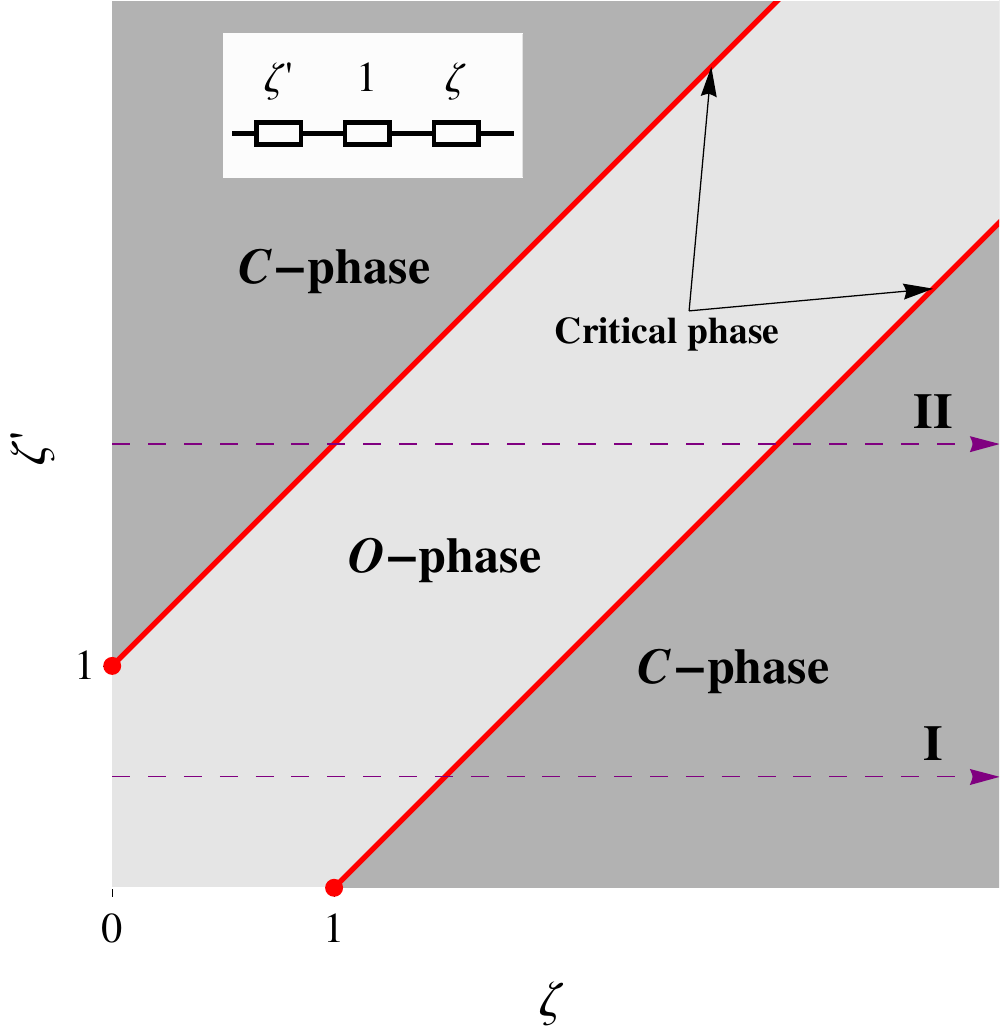}
 \caption{Main panel: the phase diagram of DTE. Red lines correspond to the critical phase ($
 |\zeta-\zeta'|=1$). Dashed line I (II) demonstrates the single (double) DTE transition when one edge reflection (i.e., $\zeta'$) is fixed and the second (i.e., $\zeta$) increases. Inset: Macroscopic interpretation of phase parameters $\zeta,\zeta'$. The total resistance (inverse average conductance) of the sample is the sum of the left and right edge resistance and the bulk resistance. They are $\zeta'$, $\zeta$ and $1$, respectively when rescaled by the bulk resistance $L/\xi$.}
  \label{fig:8}
\end{figure}

\subsection{Phase structure of DTE}
\label{sec:two_kinds_of_phase_structures}

We note that compared to the quasi $1$D sample without edge reflections, two new length scales appear in the presence of edge reflections. That is, the characteristic length $z_b$ ($z_b'$) associates with the reflection of the right (left) sample edge. More precisely, the ratio of this length to the transport mean free path, $\ell$, depends only on the reflection of the corresponding sample edge and monotonically increases with this edge reflection: it is of order unity when edge reflection vanishes and diverges when edge reflection reaches unity. The explicit expression of this ratio has been obtained by various methods over several decades \cite{Rossum99} and, most recently, by the field theoretic approach \cite{Tian08}. Because it is not essential to the present work, we shall not discuss this further.

By using first-principles field theory, we find that the DTE or, equivalently, the factor
\begin{equation}\label{eq:53}
    f({\cal T})\equiv \frac{\rho({\cal T})}{\rho_{0}({\cal T})}
\end{equation}
characterizing the deviation from the bimodal distribution $\rho_0({\cal T})$, depends on only two dimensionless parameters, $\zeta = z_b / L , \zeta' = z_b'/L$. The physical meaning of these two parameters will become clearer in Sec.~\ref{sec:physical_meaning_parameter}. Here we emphasize that the details of sample structure (e.g., disorder configuration, edge reflection, etc.) only affect their values as well as the average bulk conductance $\xi/L$, namely, the conductance of the sample without the end reflections. The parameter of $\xi/L$ is irrelevant since it is an overall factor of $\rho_0({\cal T})$. So the behavior of DTE is universal. For $|\zeta-\zeta'|<1$, the high transmission eigenchannels are opened giving rise to a singularity $\sim (1-{\cal T})^{-\frac{1}{2}}$ as ${\cal T}\rightarrow 1$. For $|\zeta-\zeta'|=1$, although the high transmission eigenchannels are still present, a distinct singularity $\sim (1-{\cal T})^{-\frac{1}{3}}$ occurs. For $|\zeta-\zeta'|>1$, the high transmission eigenchannels are blocked; correspondingly, the asymptotic behavior of DTE undergoes a drastic change: $\rho({\cal T}\rightarrow 1)=0$ and consequently $\rho({\cal T})$ becomes unimodal (with the peak near zero). These asymptotic behaviors as ${\cal T}\rightarrow 1$ do not depend on the specific values of $\zeta,\zeta'$, rather on whether the asymmetry parameter $|\zeta-\zeta'|$ is smaller (larger) than or equal to unity. This feature allows one to define three phases -- $O$, $C$, and Critical -- where the DTE behaves in qualitatively different ways (see Table~\ref{Table3} for a summary), resulting in the $\zeta$-$\zeta'$ phase diagram (Fig.~\ref{fig:8}). It is symmetric with respect to the line of $\zeta=\zeta'$, reflecting the invariance of the DTE with respect to the exchange of the two edge reflections.

Whenever a path in this phase diagram crosses a critical line, the high transmission eigenchannels are switched on or off. As such, tuning the values of $\zeta,\zeta'$ leads to rich phase transition phenomena. To show this, we consider a path corresponding to fixing the reflection of one edge (say, the left, i.e., $\zeta'$) while increasing that at the other (i.e., $\zeta$). We find that the DTE exhibits either single or double transitions: single transition occurs for $\zeta'<1$ (Line I in Fig.~\ref{fig:8}); double transitions occur for $\zeta'>1$ (Line II). We see that the previous result \cite{Tian13} corresponds to the special line of $\zeta'=0$ (or $\zeta=0$) in this phase diagram. Importantly, the stripe-like regime corresponding to the $O$-phase extends to infinity. This implies that the high transmission eigenchannels can still exist even when edge reflections are large, provided that they are not strongly asymmetric. This phenomenon resembles resonant transmission and signals strong interference origin of the high transmission eigenchannel. We shall explore this below.

\subsection{Physical meaning of parameters $\zeta,\zeta'$}
\label{sec:physical_meaning_parameter}

To better understand the physical meaning of controlled parameters $\zeta,\zeta'$ we consider the average conductance $g$ defined as
\begin{equation}\label{eq:51}
    g=\int_0^1 d{\cal T}{\cal T}\rho({\cal T}).
\end{equation}
From the analytical theory developed below for $\rho({\cal T})$ it follows that, for arbitrary strength of edge reflections,
\begin{equation}\label{eq:52}
    g=\frac{\xi}{L+z_b+z_b'}.
\end{equation}
We are not aware of any exact derivations of this result in the literature for arbitrary values of $z_b/L$ and $z_b'/L$, although Eq.~(\ref{eq:52}) has been used in Ref.~\onlinecite{Genack12}.
Equation (\ref{eq:52}) gives Ohm's law (inset of Fig.~\ref{fig:8}): the reflection of the left (right) edge introduces an edge resistance $z_b'/\xi$ ($z_b/\xi$) in series with the bulk resistance (namely the resistance of the sample without edge reflections) $L/\xi$. This shows that, in sharp contrast to DTE, $g$ does not exhibit any criticality. Moreover, it shows that $\zeta,\zeta'$ are the ratios of the corresponding edge resistance to the bulk resistance.

Therefore, the role of these two parameters is two-fold: at the macroscopic level (i.e., as macroscopic observables such as the average conductance are concerned), they are essentially the edge resistors in series with the bulk one; at the mesoscopic level (i.e., as mesoscopic quantities such as the DTE are concerned), they play the role of phase parameters.

\subsection{Physical picture}
\label{sec:physical_mechanism}

\begin{figure}[h]
  \centering
\includegraphics[width=8.0cm]{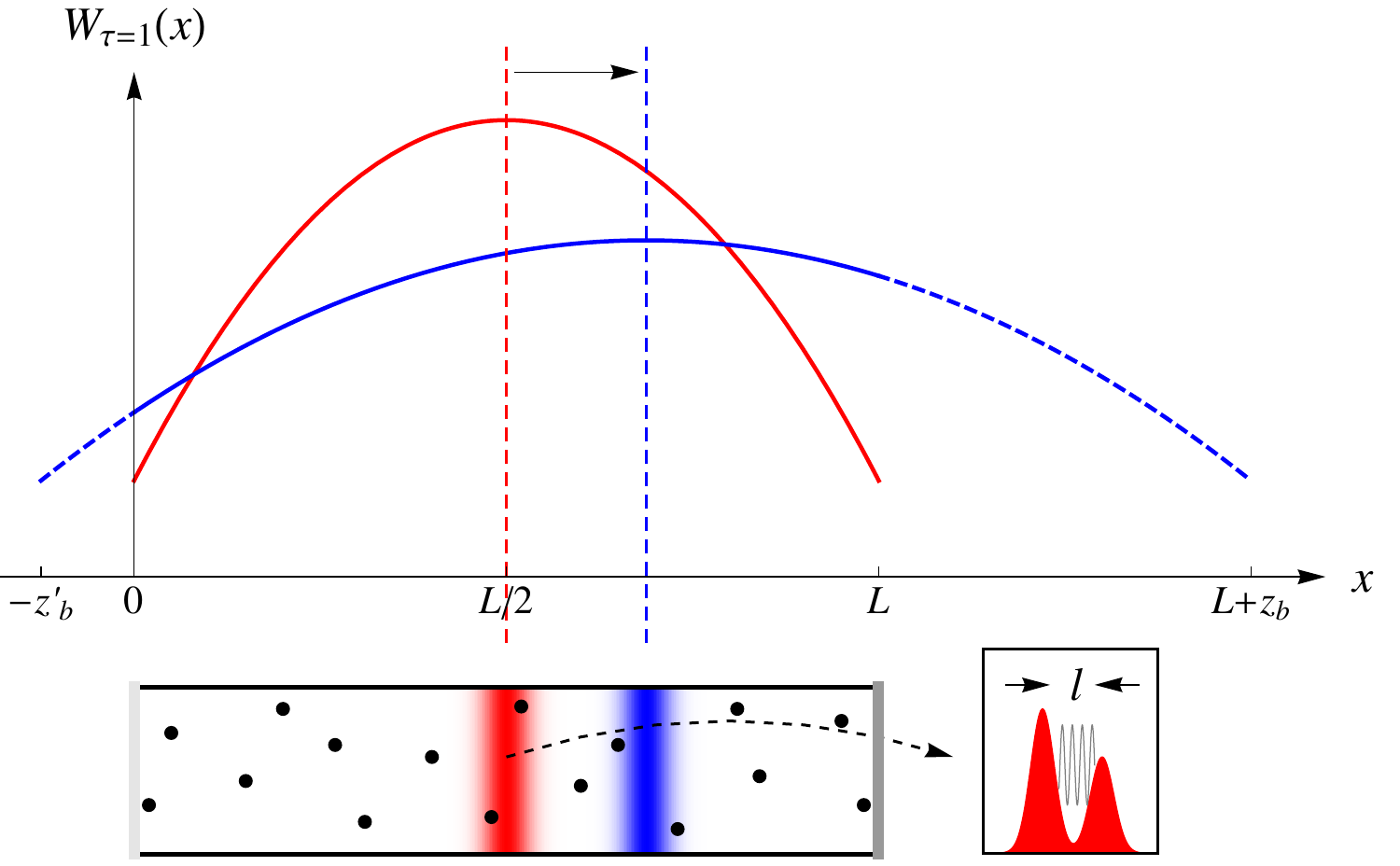}
 \caption{Effective resonator related to a transmission eigenchannel with ${\cal T}=1$ is located in the center of the diffusive sample (left lower panel, red area) and is of the size of the transport mean free path $\ell$ (right lower panel).  This resonator is bound to the center of the corresponding eigenchannel profile $W_{{\cal T}=1}(x)$ (upper panel, red curve). When the reflections of the sample edges are asymmetric, the center of the profile is shifted towards the edge with stronger reflection (upper panel, blue curve), and so is the resonator center (left lower panel, blue area).}
  \label{fig:9}
\end{figure}

The transmission eigenchannel is associated with an energy profile (integrated over the transverse coordinate) across the sample. This profile is denoted as $W_{{\cal T}}(x)$, with $x$ being the distance to the left sample edge and ${\cal T}$ the corresponding eigenvalue. For perfectly transmitting eigenchannels (${\cal T}=1$) in diffusive samples without edge reflections, this profile is a parabola with its center at $x=L/2$ (Fig.~\ref{fig:9}, red curve in the upper panel) \cite{Genack15}. In the presence of edge reflections, this profile remains parabolic but its center moves to the side with stronger edge reflection, i.e., larger edge resistance (Fig.~\ref{fig:9}, blue curve in the upper panel). As we will see later, analytical and numerical analyses suggest that, strikingly, there is a $1$D resonator bound to the profile $W_{{\cal T}=1}(x)$ (Fig.~\ref{fig:9}, lower panel). This resonator is small, with a size of order of the transport mean free path, $\ell$. It is formed by disordered barriers. It is analogous to an effective cavity in which the states are trapped in the regime of $1$D strong localization \cite{Freilikher03}. Importantly, it gives rise to perfect transmission (but not to the eigenchannel profile which is essentially the probability density for a diffusive wave to return to a cross section at depth $x$). If the asymmetry in edge reflections is increased continuously, the resonator moves together with the center of profile $W_{{\cal T}=1}(x)$ to the sample edge, and eventually the DTE transition occurs.

\section{Analytical theory of DTE}
\label{sec:transition_diffusive}

In this section we study analytically impacts of edge reflection on the DTE. An essential difference from the earlier work \cite{Tian13} is that in Ref.~\onlinecite{Tian13} the reflection appears only at one sample edge while in the present work both edges have finite reflectivities. As we show below, this difference introduces even more interesting physical phenomena. The generalization of the earlier analytical approach to the present problem is, however, a highly nontrivial task.


\subsection{General formalism}
\label{sec:formalism}

We adopt the same general formalism as that of Ref.~\onlinecite{Tian13}, with an importance difference that we outline below. Our starting point is an exact expression for the DTE,
\begin{eqnarray}
\rho({\cal T}) &=& \frac{1}{2\pi}[F(\phi+i\pi)+F^*(\phi+i\pi)]
\frac{d\phi}{d{\cal T}}, \label{eq:1}\\
F(\phi)&\equiv& -\frac{i}{2}\sinh\phi\left\langle
{\rm tr}\left[\frac{{\bf t}{\bf t}^\dagger}
{1+\sinh^2(\phi/2){\bf t}{\bf t}^\dagger}\right]\right\rangle,
\label{eq:2}
\end{eqnarray}
where $\phi$ is understood as $\phi-i\delta$
with $\delta$ a positive infinitesimal. The transmission eigenvalue ${\cal T}$ is related to the parameter $\phi$ through
\begin{equation}\label{eq:33}
    {\cal T}\equiv \cosh^{-2}(\phi/2).
\end{equation}
Then, it is a canonical method of casting the function, $F(\phi)$, into a functional integral over the supersymmetric
field $Q(x)$, where $Q \equiv\{Q_{\alpha\alpha'}^{\lambda\lambda'}\}
$ is a $4\times 4$ supermatrix, with $\lambda,\lambda'=1,2$ denoting the advanced-retarded (`ar') sector and $\alpha,\alpha'={\rm f,b}$ the fermionic-bosonic (`fb') sector \cite{Efetov97}. Because the results in this work do not depend on whether the time-reversal symmetry is present, we consider the system with broken time-reversal symmetry. This gives rise to a $4\times 4$ supermatrix structure; otherwise one has to introduce additional matrix index to accommodate the time-reversal symmetry. The result reads
\begin{eqnarray}
F(\phi)&=& -\frac{i\xi}{2}
\int_{(2z_b' Q\partial_x Q+[Q,\Lambda])|_{x=0}=0}^{(2z_b Q\partial_x Q - [Q,\Gamma])|_{x=L}=0} D[Q]\nonumber\\
&& \times (Q\partial_x Q)^{21}_{\rm bb}|_{
x=0,\theta=i\phi} e^{-\frac{\xi}{8}
\int_0^L dx
{\rm str} (\partial_x Q)^2},
\label{eq:3}
\end{eqnarray}
with `str' being the supertrace. Here $\Lambda$ and $\Gamma$ are constant supermatrices,
\begin{eqnarray}
  \Lambda &=& \left(
                \begin{array}{cc}
                  \mathbbm{1}^{\rm fb} & 0 \\
                  0 & -\mathbbm{1}^{\rm fb} \\
                \end{array}
              \right)^{\rm ar},\label{eq:16}\\
\Gamma &=&
                   \left(
\begin{array}{cc}
                        \cos\theta & -i\sin\theta \\
                        i\sin\theta & -\cos\theta \\
                      \end{array}
                    \right)^{\rm ar}\oplus
\left(
                      \begin{array}{cc}
                        \cosh\phi & \sinh\phi \\
                        -\sinh\phi & -\cosh\phi \\
                      \end{array}
                    \right)^{\rm ar}\,\,\label{eq:17}
\end{eqnarray}
with $0<\theta<\pi$. Most importantly, edge reflection imposes boundary constraints on the supermatrix field, $2z_b' Q\partial_x Q+[Q,\Lambda]=0$, at the left edge and $2z_b Q\partial_x Q - [Q,\Gamma]=0$ at the right. As first shown in Ref.~\onlinecite{Tian08}, these boundary constraints are the only effect of sample openness, namely the exchange of wave energy of media with external environments on the edges, on field theory constructions. This notwithstanding, the constraints take rich physical wave phenomena in open media into account (for a review, see Ref.~\onlinecite{Tian13a}).

Equations (\ref{eq:1}), (\ref{eq:2}), and (\ref{eq:3}) constitute an exact formalism for calculating the DTE. We emphasize that this formalism is very general and valid for both quasi $1$D diffusive and localized systems, although the detailed treatments are very different. Also, the values of $z_b,z_b'$ could be arbitrarily large: this advantage allows us to thoroughly explore impacts of arbitrarily strong edge reflections. In the remainder of this section we apply it to the former case where $L\ll \xi$.

\begin{figure}[h]
  \centering
\includegraphics[width=8.0cm]{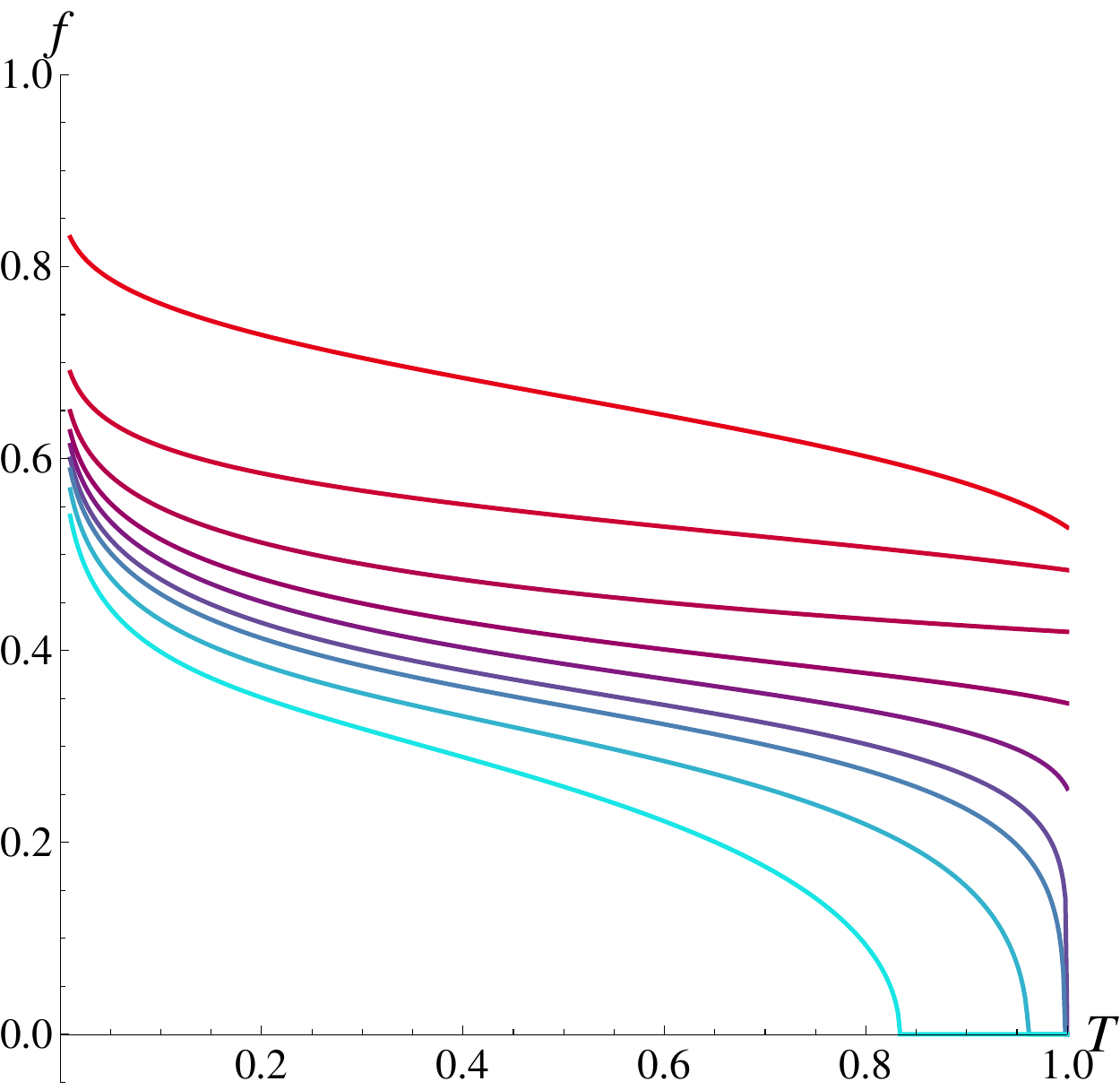}
 \caption{Deviation factor $f$ as a function of the edge reflection parameter $\zeta$ ($\zeta'$ is fixed). DTE exhibits a single transition $C \rightarrow {\rm Critical} \rightarrow O$ as $\zeta$ increases from zero and $\zeta'$ is fixed to $0.6$. The deviation factor is obtained by solving Eqs.~(\ref{eq:7}) and (\ref{eq:8}) numerically. The transition corresponds to $\zeta=1.6$ with $f({\cal T}=1)=0$ (dark blue line):  above this value the probability density of perfectly transmitting eigenchannels is fully suppressed. From top to bottom the value of $\zeta$ is $0,0.3,0.6,0.9,1.2,1.6,2,3$, and $5$.}
  \label{fig:1}
\end{figure}

\begin{table*}
\newcommand{\tabincell}[2]
{\begin{tabular}{@{}#1@{}}#2
\end{tabular}}
\centering
\caption{\label{Table1} Numerical values of $\alpha$ and $\beta$ obtained by numerically solving Eqs.~(\ref{eq:7}) and (\ref{eq:8}) at $\zeta-\zeta'=1$.}
\begin{tabular}{ccccccccc}
  \hline\hline
  $\zeta$ & $1$ & $3$ & $5$ & $7$ & $9$ & $11$ & $13$ & $15$ \\
  \hline
  $\alpha$ & $0.333331$\, & $0.333228$\, & $0.333078$\, & $0.332906$\, & $0.332716$\, & $0.332510$\, & $0.332292$\, & $0.332062$\, \\
  \hline
  $\beta$ & $0.333834$\, & $0.339621$\, & $0.342927$\, & $0.345654$\, & $0.348075$\, & $0.350291$\, & $0.352356$\, & $0.354303$\, \\
  \hline\hline
\end{tabular}
\end{table*}

\subsection{Saddle point configurations}
\label{sec:saddle_point}

Because of $\xi/L\gg 1$ the functional integral over the $Q$-field is dominated
by fluctuations around the saddle point configurations
which satisfy
\begin{equation}\label{eq:54}
    \partial_x (Q\partial_xQ)=0
\end{equation}
and the boundary conditions
\begin{eqnarray}\label{eq:55}
    (2z_b' Q\partial_x Q+[Q,\Lambda])|_{x=0}=0,\nonumber\\
    (2z_b Q\partial_x Q - [Q,\Gamma])|_{x=L}=0.
\end{eqnarray}
The solution to the saddle point equation (\ref{eq:54}) has the same structure as $\Gamma$ given by Eq.~(\ref{eq:17}), i.e.,
\begin{eqnarray}
Q(x)&=&
                   \left(
\begin{array}{cc}
                        \cos\Theta(x) & -i\sin\Theta(x) \\
                        i\sin\Theta(x) & -\cos\Theta(x) \\
                      \end{array}
                    \right)^{\rm ar}\nonumber\\
                    &\oplus&
\left(
                      \begin{array}{cc}
                        \cosh\Phi(x) & \sinh\Phi(x) \\
                        -\sinh\Phi(x) & -\cosh\Phi(x) \\
                      \end{array}
                    \right)^{\rm ar}.\,\,\label{eq:18}
\end{eqnarray}
With the substitution of Eq.~(\ref{eq:18}), the saddle point equation (\ref{eq:54}) is reduced to
\begin{equation}
\partial_x^2 \Phi = \partial_x^2 \Theta = 0,
\label{eq:4}
\end{equation}
and the boundary conditions (\ref{eq:55}) to
\begin{eqnarray}\label{eq:2a}
    &&\left(z_b' \partial_x \Phi - \sinh\Phi\right)|_{x=0} \nonumber\\
    &=&\left(z_b' \partial_x \Theta - \sin\Theta\right)|_{x=0}=0
\end{eqnarray}
and
\begin{eqnarray}
\label{eq:2b}
&&\left(z_b \partial_x \Phi + \sinh(\Phi - \phi)\right)|_{x=L}\nonumber\\
&=&\left(z_b \partial_x \Theta + \sin(\Theta-\theta)\right)|_{x=L}=0,
\end{eqnarray}
respectively.

The solution to Eq.~(\ref{eq:4}) has the general form of
\begin{eqnarray}
  \Phi(x)&=&C_\phi x/L + \phi_0, \label{eq:56}\\
  \Theta (x)&=&C_\theta x/L + \theta_0, \label{eq:57}
\end{eqnarray}
where the coefficients $C_{\phi,\theta}\in \mathbbm{R}$,
$\phi_0>0$ and $0<\theta_0<\pi$. Due to the compactness of the fermionic component, i.e., the $2\pi$-periodicity of the sine and cosine functions, there are a family of saddle point solutions, each of which corresponds to a saddle point action proportional to $\frac{\xi}{L}(C_{\theta=i\phi}^2+C_\phi^2)$. Assuming that the compactness does not play any role, we may carry out
the analytic continuation of the second equations in the boundary constraints (\ref{eq:2a}) and (\ref{eq:2b}), which gives $C_{\theta=i\phi}=iC_\phi$.
This is the saddle point configuration with the smallest action which is zero. It preserves the supersymmetry. The other saddle point configurations arise from the compactness and break the supersymmetry. They are
gapped by an action $\sim{\cal O}(\frac{\xi}{L})$. This action is large for diffusive waves and the supersymmetry broken saddle points are can thereby be neglected. Furthermore, we may
ignore fluctuations around the saddle point in the pre-exponential factor of Eq.~(\ref{eq:3})
since they only give rise to corrections of lower order. Finally, by integrating out
Gaussian fluctuations around the supersymmetric saddle point, which gives a functional superdeterminant of
unity at $\theta=i\phi$ because of the supersymmetry, we reduce Eq.~(\ref{eq:3}) to
\begin{equation}
    F(\phi)=-\frac{i}{2}\frac{\xi}{L}C_\phi.
\label{eq:19}
\end{equation}
This is uniquely determined by the coefficient $C_\phi$ and we therefore focus on the solution of $\Phi(x)$ below.

\begin{figure}
\centering
\begin{minipage}{0.48\linewidth}
  \centerline{\includegraphics[width=7.0cm]{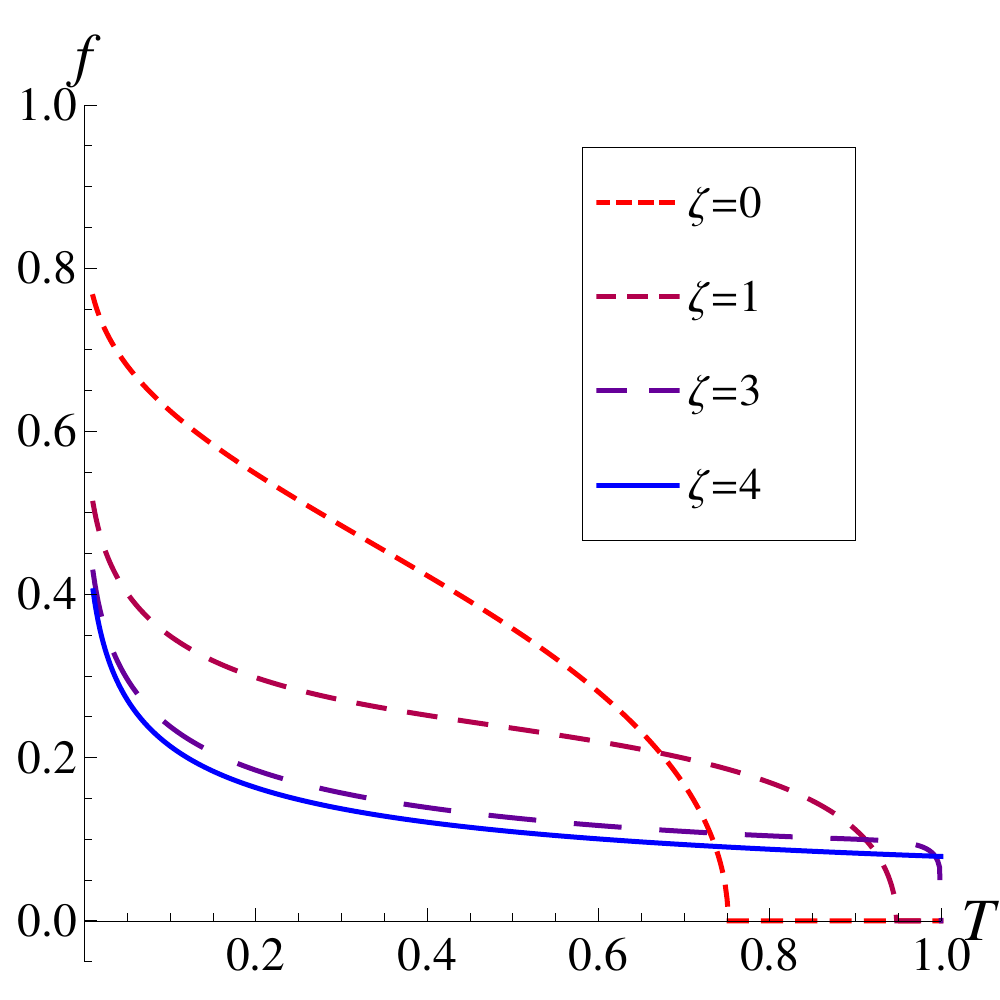}}
\vfill
  \centerline{\includegraphics[width=7.0cm]{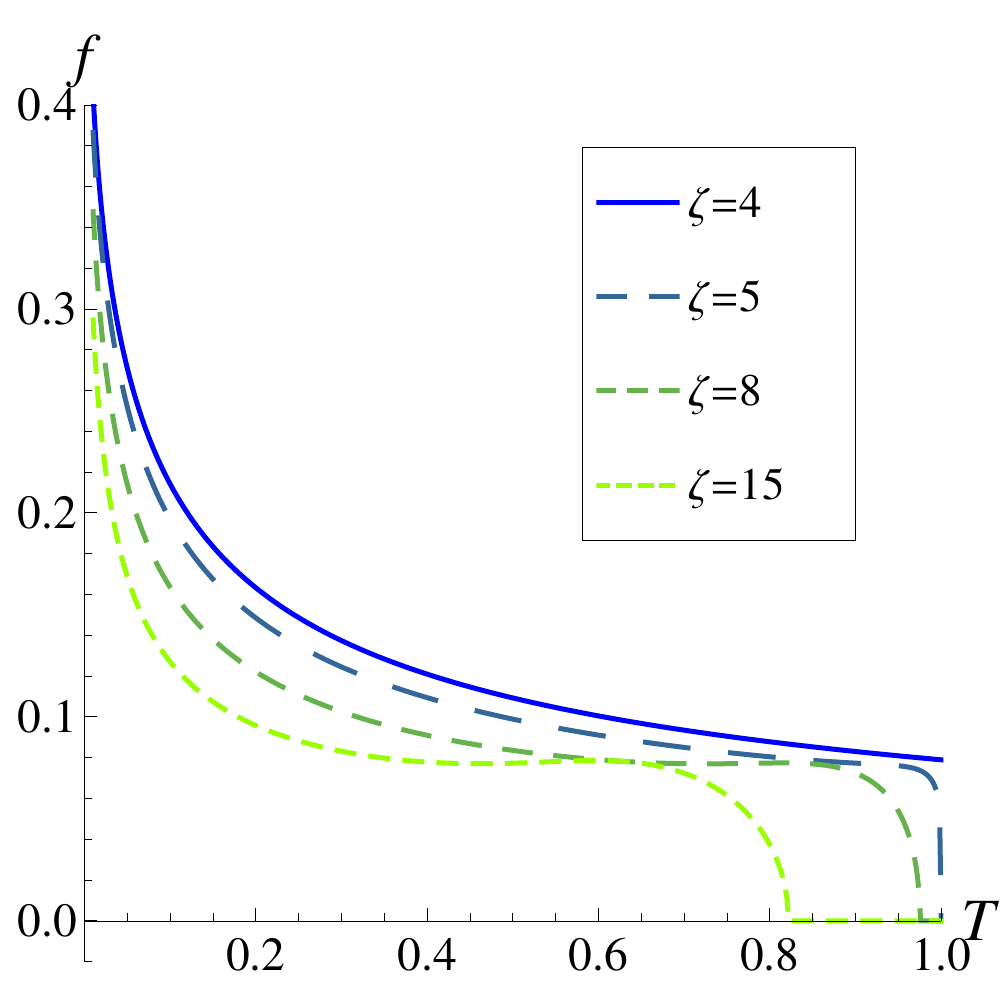}}
\end{minipage}
\caption{Deviation factor $f$ as a function of the edge reflection parameter $\zeta$ ($\zeta'$ is fixed). DTE exhibits double transitions $C \rightarrow {\rm Critical} \rightarrow O \rightarrow {\rm Critical} \rightarrow C$ when $\zeta$ increases from zero and $\zeta'$ is fixed to $4$. The deviation factor is obtained by solving Eqs.~(\ref{eq:7}) and (\ref{eq:8}) numerically. The transitions correspond to $\zeta=3$ (upper panel, purple dashed line) and $\zeta=5$ (lower panel, dark green dashed line), respectively. $f({\cal T}=1)$ is positive for $3<\zeta<5$ and otherwise zero.}
\label{fig:2}
\end{figure}

For the limiting case of $z_b,z_b'=0$, the boundary constraints (\ref{eq:2a}) and (\ref{eq:2b}) [for $\Phi(x)$] are reduced to
\begin{eqnarray}\label{eq:46}
    \Phi|_{x=0}=0,\quad
\Phi|_{x=L}=\phi.
\end{eqnarray}
Combined with Eq.~(\ref{eq:56}) they give $\Phi(x)=\phi x/L$. Substituting it into
Eqs.~(\ref{eq:1}) and (\ref{eq:19}) gives Eq.~(\ref{eq:20}).

By using Eqs.~(\ref{eq:1}) and (\ref{eq:19}) and taking into account the parametrization of ${\cal T}$ we obtain
\begin{equation}\label{eq:3f}
    f(\phi)=(C_{\phi+i\pi}-C_{\phi-i\pi})/(2i\pi)\equiv \Delta C_\phi/(2i\pi).
\end{equation}
According to this equation we need in principle to find $C_\phi$ first and then make analytic continuations. But this is a difficult task because as we will see shortly, the equation satisfied by $C_\phi$ is transcendental. To overcome this difficulty we will establish a closed equation of $f(\phi)$ below and solve the equation.

\subsection{DTE transition}
\label{sec:transition_theory}

From Eqs.~(\ref{eq:2a}) and (\ref{eq:2b}) we find
\begin{eqnarray}
  && \left(\zeta' C_\phi-\sinh\phi_0\right)|_{x=0} = 0,\label{eq:3a}\\
  && \left(\zeta C_\phi + \sinh(C_\phi + \phi_0 - \phi)\right)|_{x=L} = 0.\label{eq:3b}
\end{eqnarray}
Substitution Eq.~(\ref{eq:3a}) into Eq.~(\ref{eq:3b}) gives
\begin{equation}
\cosh\phi_0 = -\dfrac{\zeta' \cosh(C_\phi - \phi)C_\phi + \zeta C_\phi}{\sinh(C_\phi - \phi)}.
\label{eq:3c}
\end{equation}
Combining Eqs.~(\ref{eq:3a}) and (\ref{eq:3c}) we find
\begin{equation}
\dfrac{[\zeta' \cosh(C_\phi - \phi)C_\phi + \zeta C_\phi]^2}{\sinh^2(C_\phi-\phi)}-\zeta'^2 C_\phi^2 = 1.
\label{eq:3d}
\end{equation}
This equation may be rewritten as
\begin{equation}
C_\phi^2 = \dfrac{\sinh^2(C_\phi - \phi)}{2\zeta \zeta' \cosh(C_\phi - \phi) +\zeta^2 + \zeta'^2},
\label{eq:3e}
\end{equation}
which implicitly gives $C_\phi$ as a function of $\phi$, $\zeta$, and $\zeta'$.

We introduce $\overline{C}_\phi \equiv (C_{\phi+i\pi} + C_{\phi-i\pi})/2$. Then, by using Eq.~(\ref{eq:3e}) we obtain
\begin{eqnarray}
2 \overline{C}_\phi^2 +\Delta C_\phi^2/2 =
\dfrac{\sinh^2 \psi_+}{a \cosh\psi_+ +b} + \dfrac{\sinh^2 \psi_-}{a \cosh\psi_- +b}
\label{eq:4a}
\end{eqnarray}
and
\begin{eqnarray}
2 \overline{C}_\phi \Delta C_\phi &=&
\dfrac{\sinh^2 \psi_+}{a \cosh\psi_+ +b} - \dfrac{\sinh^2 \psi_-}{a \cosh\psi_- +b},
\label{eq:4b}
\end{eqnarray}
where $\psi_\pm \equiv C_{\phi\pm i\pi}-(\phi\pm i\pi)$, $a = 2\zeta \zeta'$, and $b = \zeta^2 + \zeta'^2$.
Equations (\ref{eq:4a}) and (\ref{eq:4b}) are equivalent to (see Appendix~\ref{sec:calculations} for the derivation)
\begin{widetext}
\begin{equation}\label{eq:7}
2 (\overline{C}_\phi^2 -\pi^2 f^2) =\dfrac{-a(\cosh2({\overline C}_\phi-\phi) +\cos 2\pi f-2)\cosh({\overline C}_\phi-\phi)\cos \pi f+b(\cosh2({\overline C}_\phi-\phi)\cos 2\pi f-1)}
{a^2(\cosh2({\overline C}_\phi-\phi)+\cos 2\pi f)/2-2ab\cosh ({\overline C}_\phi-\phi)\cos \pi f+b^2},
\end{equation}
\begin{equation}\label{eq:8}
4\pi \overline{C}_\phi f =\dfrac{-a(\cosh2({\overline C}_\phi-\phi) +\cos 2\pi f+2)\sinh({\overline C}_\phi-\phi)\sin\pi f+b\sinh2({\overline C}_\phi-\phi)\sin 2\pi f}
{a^2(\cosh2({\overline C}_\phi-\phi)+\cos 2\pi f)/2-2ab\cosh({\overline C}_\phi-\phi)\cos \pi f+b^2}.
\end{equation}
\end{widetext}
Recall that ${\overline C}_\phi$ is real and $f>0$. Equations (\ref{eq:7}) and (\ref{eq:8}) constitute the closed equations for $\overline{C}_\phi$ and $f$. Giving $\zeta,\zeta'$ we may solve them numerically and find the deviation factor $f$. The representative results of $f$ are shown in Figs.~\ref{fig:1} and \ref{fig:2}.

As shown in Fig.~\ref{fig:1}, if the reflection of the left edge, i.e., $\zeta'$, is fixed and sufficiently small, upon increasing the reflection of the right edge, i.e., $\zeta$, the DTE exhibits a single transition at certain critical value of $\zeta$ with a hallmark of $f({\cal T}=1)=0$: below the critical value the maximum transmission eigenvalue is smaller than unity while above the critical value it is unity. This transition is similar to the previous result \cite{Tian13} where $\zeta'$ is set to zero. As shown in that work, in such limiting case the transition coincides (but generally not) with a DTE transition predicted for a completely different setup, i.e., a normal metal with a single barrier placed inside the sample \cite{Nazarov94}.

The behavior is even more interesting when the fixed $\zeta'$ is large. (For this case we are not aware of any analogs of the results derived below in other wave systems.) As shown in Fig.~\ref{fig:2}, upon increasing $\zeta$, the DTE exhibits double transitions at two critical values of $\zeta$: below (above) the lower (upper) critical value the maximum transmission eigenvalue is smaller than unity while between these two critical values is unity; the critical points have a hallmark of $f({\cal T}=1)=0$.

\subsection{Exact criterion for DTE transition}
\label{sec:criterion}

Next, we derive the exact criterion for the transition. Because a hallmark of
the latter is $f({\cal T}=1)=0$, as mentioned above, we set $\phi=0$
[recalling that ${\cal T}=\cosh^{-2}(\phi/2)$] and $f=0$ in
Eqs.~(\ref{eq:7}) and (\ref{eq:8}). We find that the latter equation is satisfied automatically and the former is reduced to
\begin{eqnarray}\label{eq:9}
2 \overline{C}_\phi^2
=\dfrac{\cosh2{\overline C}_\phi-1}
{b-a\cosh {\overline C}_\phi}.
\end{eqnarray}
To find the condition under which this equation has a real solution of $\overline{C}_\phi>0$, we rewrite Eq.~(\ref{eq:9}) as
\begin{equation}\label{eq:10}
F(\overline{C}_\phi)\equiv 2\overline{C}_\phi^2(b-a\cosh\overline{C}_\phi) - (\cosh2\overline{C}_\phi-1)=0.
\end{equation}
Performing the Taylor expansion for $F(\overline{C}_\phi)$ gives
\begin{equation}
\label{eq:11}
F(\overline{C}_\phi)=2(b-a-1)\overline{C}_\phi^2-\sum_{n=2}^\infty \left[\dfrac{2a}{(2(n-1))!}+\dfrac{2^{2n}}{(2n)!}\right]\overline{C}_\phi^{2n}
\end{equation}
for $|\overline{C}_\phi|<\infty$. From this we see that provided
\begin{equation}\label{eq:12}
b-a-1>0 \Rightarrow |\zeta-\zeta'|>1,
\end{equation}
$F(\overline{C}_\phi)$ is a non-monotonic function of $\overline{C}_\phi$: it first increases from zero, then decreases and eventually decays as $\sim -e^{2\overline{C}_\phi}$ when $\overline{C}_\phi$ is sufficiently large. In this case, Eq.~(\ref{eq:10}) must have a positive root. If the inequality (\ref{eq:12}) is not satisfied, then $F(\overline{C}_\phi)$ monotonically decreases from zero and Eq.~(\ref{eq:10}) has only a trivial solution of $\overline{C}_\phi=0$ and, as a result, $f({\cal T}=1)$ must not vanish.

The inequality (\ref{eq:12}) defines the two regimes of $C$-phase in Fig.~\ref{fig:8}. It implies that the perfectly transmitting eigenchannel is blocked only if edge reflections are highly asymmetric. Otherwise, $|\zeta-\zeta'|<1$ which defines the regime of $O$-phase in Fig.~\ref{fig:8}. The transition occurs at
\begin{equation}\label{eq:21}
|\zeta-\zeta'|=1.
\end{equation}
This gives the two critical lines in Fig.~\ref{fig:8}.
From this we see that for $\zeta'<1$ single transition occurs upon increasing $\zeta$ from zero, and the critical point is ($\zeta'+1$). This phase structure is represented by Line I in Fig.~\ref{fig:8}. In the particular case of $\zeta'=0$ this result is in agreement with that found in the previous work \cite{Tian13}. For $\zeta'>1$ two transitions occur upon increasing $\zeta$ from zero, and the critical points are ($\zeta'\pm 1$). This phase structure is represented by Line II in Fig.~\ref{fig:8}.

\subsection{Criticality of DTE transition}
\label{sec:discussions}

\subsubsection{Critical statistics}
\label{sec:critical_scaling}

\begin{figure}[h]
  \centering
\includegraphics[width=8.0cm]{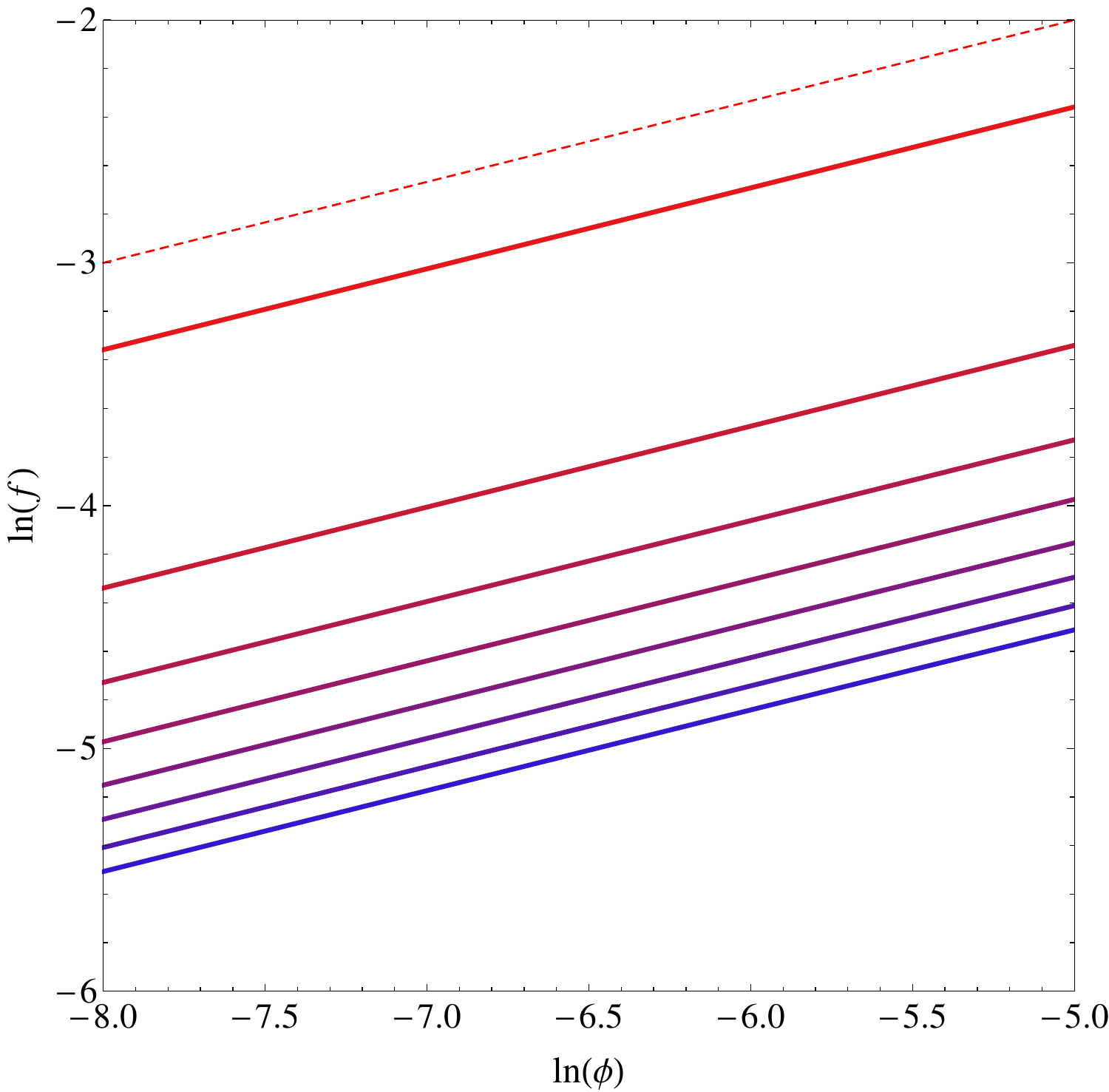}
 \caption{The numerical solution of $f(\phi)$ for different values of $\zeta$ and $\zeta'$ satisfying $\zeta-\zeta'=1$ (solid lines). From top to bottom $\zeta=1,3,5,7,9,11,13,15$. To guide the eye we also plot a dashed line with a slope of $1/3$.}
  \label{fig:3}
\end{figure}

From the results above we find that the DTE peak near ${\cal T}=1$ diverges as
\begin{equation}\label{eq:23}
    \rho({\cal T}\rightarrow 1)\sim(1-{\cal T})^{-\frac{1}{2}},\quad |\zeta-\zeta'|<1,
\end{equation}
since $f({\cal T}=1)\neq 0$. In this part we study the asymptotic behavior of $\rho({\cal T}\rightarrow 1)$ at the critical line. To this end we solve Eqs.~(\ref{eq:7}) and (\ref{eq:8}) numerically. We find that for $\phi\rightarrow 0$ corresponding to ${\cal T}\rightarrow 1$ the solution has the general form as follows (cf. Fig.~\ref{fig:3}),
\begin{equation}\label{eq:24}
    f\propto \phi^\alpha,\quad {\overline C}_\phi-\phi\propto \phi^\beta.
\end{equation}
The exponents, $\alpha,\beta$, for different values of $\zeta,\zeta'$ are given in Table \ref{Table1}. (Because both the values $\zeta'-\zeta=\pm 1$ lead to the same results, without loss of generality we consider $\zeta-\zeta'=1$.) These numerical values give $\alpha\approx \beta\approx 1/3$. In Appendix \ref{sec:expansion}, we show that a stronger relation, i.e.,
\begin{equation}\label{eq:25}
    \alpha=\beta=1/3
\end{equation}
exists. Then, combining Eqs.~(\ref{eq:24}) and (\ref{eq:25}) and taking into account that $\phi\sim (1-{\cal T})^{1/2}$ for $\phi\rightarrow 0$, we find
\begin{equation}\label{eq:26}
    \rho({\cal T}\rightarrow 1)\sim(1-{\cal T})^{-\frac{1}{3}},\quad |\zeta-\zeta'|=1.
\end{equation}
We see that the divergence of the DTE peak around ${\cal T}=1$ changes whenever the critical line is crossed, associated with the emergence or disappearance of high transmission eigenchannels. In the special case of vanishing $\zeta$ or $\zeta'$, such a change in the power of the divergence from $1/2$ to $1/3$, has been found previously \cite{Tian13}.

\subsubsection{Average conductance}
\label{sec:average_conductance}

\begin{figure}
 \begin{center}
 \includegraphics[width=8.0cm]{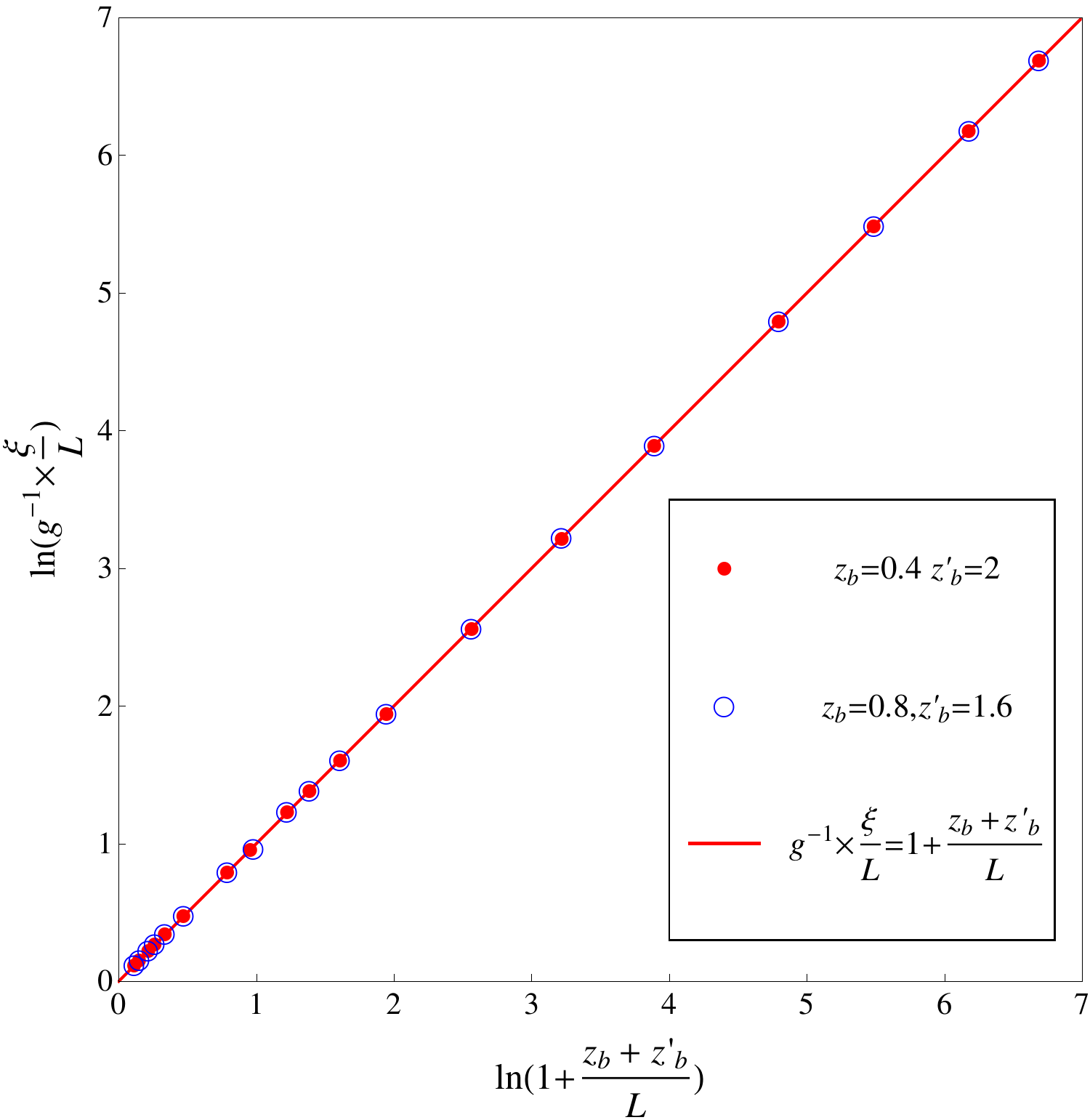}
\end{center}
 \caption{Numerically calculated Eq.~(\ref{eq:22}) (dots and circles) shows that the average conductance obeys Ohm's law irrespective of the values of $\zeta$ and $\zeta'$. Note that the two data sets corresponding to $z_b=0.4,z_b'=2$ and $z_b=0.8,z_b'=1.6$, respectively, are identical because they have the same value of $z_b+z_b'$. The slope of the fitting solid red line is one.}
 \label{fig:Ohm}
\end{figure}

Substituting $\rho({\cal T})=f({\cal T})\rho_0({\cal T})$ into the definition (\ref{eq:51}), we find that the average conductance, upon being rescaled by $\xi/L$, is given by
\begin{equation}\label{eq:22}
    g\times(L/\xi)=
    \frac{1}{2}\int_0^1 d{\cal T} \frac{f({\cal T})}{\sqrt{1-{\cal T}}}.
\end{equation}
It depends only on the parameters $\zeta$ and $\zeta'$. An interesting question is, does this observable exhibit any criticality?
Because we cannot analytically calculate Eq.~(\ref{eq:22}), we resort to carrying out the integration numerically. To this end we let $\zeta,\zeta'$ vary over a wide range from $10^{-1}$ to $10^3$. As shown in Fig.~\ref{fig:Ohm}, the numerical results are well fitted by $1/(1+\zeta+\zeta')$.
Therefore, the average conductance is given by Eq.~(\ref{eq:52}), which is none but Ohm's law and shows that the average conductance does not exhibit criticality. It agrees with the well-known result \cite{Lagendijk89,Zhu91,Genack93} obtained by using
the diffusion model or radiative transfer theory \cite{Rossum99,Morse53,Chandrasekhar}. This agreement is not accidental. Rather, it is due to that the field theory, in a somehow automatical manner, captures collective modes such as the diffuson (see Ref.~\onlinecite{Tian13a} for a pedagogical review). To see this, in Appendix~\ref{sec:diffuson} we investigate the Gaussian fluctuations in the field theory (\ref{eq:3}) and find the diffuson explicitly. The result entails the canonical physical meaning of $z_b,z_b'$ namely the so-called extrapolation length \cite{Rossum99}.

\section{Numerical test of DTE transition}
\label{sec:numerical result}

In this section we put the analytic results derived in Sec.~\ref{sec:transition_diffusive} under numerical test. To this end we launch a scalar wave into a quasi $1$D disordered waveguide which is a $900\times300$ rectangular lattice and thereby locally two-dimensional. We simulate wave transport by using the standard recursive Green's function method \cite{Baranger91,MacKinnon85,Bruno05}. Specifically, we are interested in the Green's function $G(r,r')$ between grid points $r\equiv(0,y)$ and
$r\equiv(L,y')$ at the left ($x=0$) and right ($x=L$) edges, with $y$ being the transverse coordinate. The lattice constant is the inverse wave number. The squared refractive index at each site fluctuates independently around the air background value, taking values randomly from the interval $[0.7,1.3]$. The wave velocity in the air background is set to unity. To create edge reflection, we add an additional layer of thickness $2$ with uniform refractive index at the corresponding sample edge. For given values of refractive index at the left and right edges and disorder configuration, we numerically compute the transmission matrix ${\bf t}\equiv\{t_{ba}\}$ in the basis of the empty waveguide modes, $\phi_{a,b}(y)$, where the indexes $a,b$ are the labels of the left and right edges, respectively. These matrix elements are given by
\begin{equation}\label{eq:32}
    t_{ba}=\sqrt{v_b v_a} \int dy \int dy' \phi_b (y) \phi_a^* (y') G(r,r'),
\end{equation}
where $v_a$ is the group velocity of the
empty waveguide mode $a$ at the wave frequency.
For simulations the channel number is set to $100$, i.e., ${\bf t}$ is a $100\times100$ matrix. By numerical diagonalization
we find the transmission eigenvalues, $\tau_n$. Then, we repeat the same procedures for $6000$ disorder configurations.

\begin{figure}
 \begin{center}
 \includegraphics[width=8.0cm]{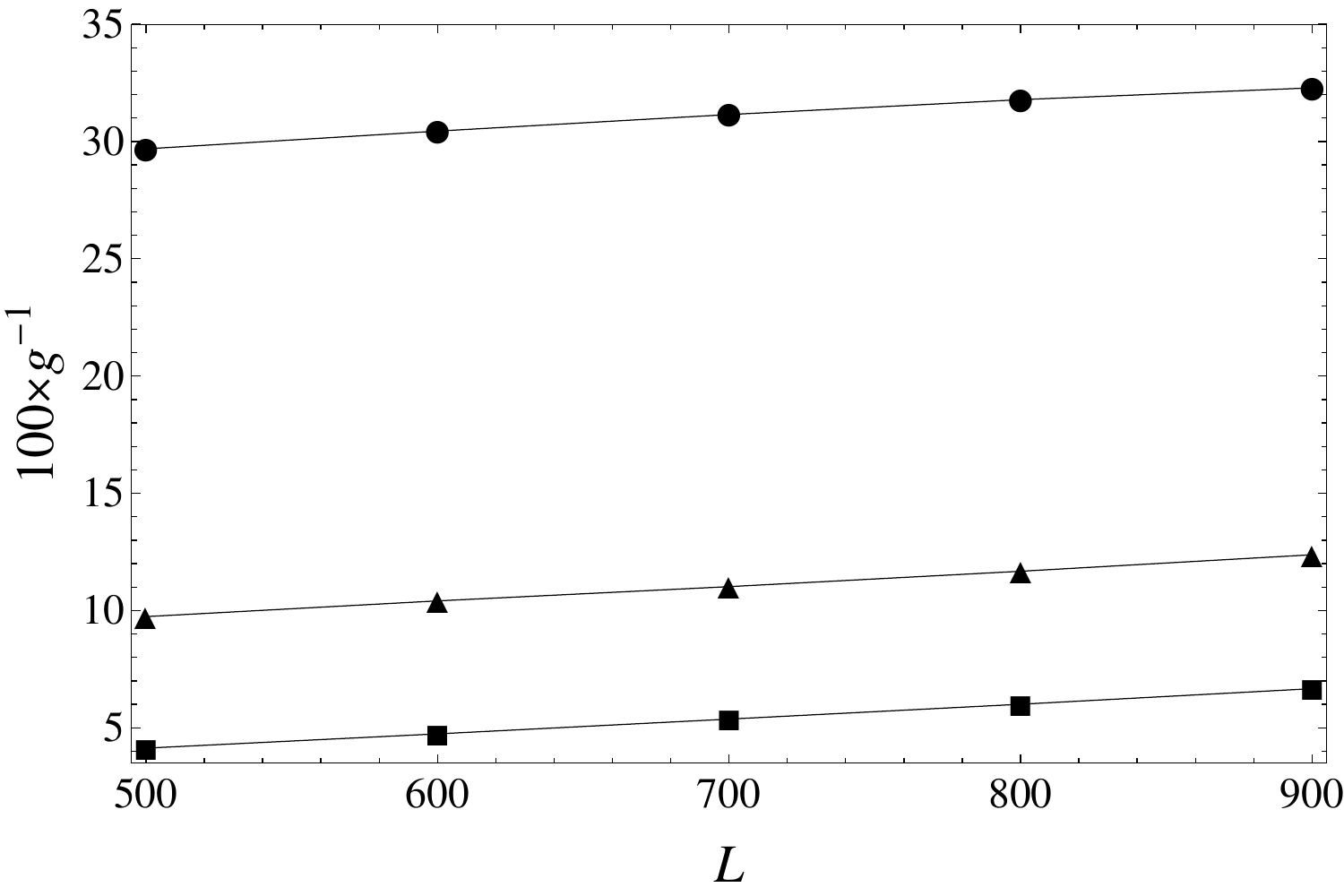}
\end{center}
 \caption{Numerical simulations show that in the absence of edge reflection the average conductance (squares) obeys Ohm's law (bottom solid line). The slope of the line gives the inverse localization length, $\xi^{-1}$, and its intersection with the axis $L=0$ ($2z_{b0}/\xi$) gives the value of $z_{b0}$. For different values of the refractive index, Ohm's law (top and middle solid lines) remains valid, with the slope of the straight lines unchanged. The refractive indexes at the left and right edges are $(1,1.9)$ (triangles) and $(1,2.1)$ (circles), respectively.}
 \label{fig:gL}
\end{figure}

\subsection{Ohm's law}
\label{sec:ohm_law}

With the transmission eigenvalues $\tau_n$ obtained by the numerical method described above, we compute the average conductance given by $g=\langle \sum_n \tau_n \rangle$. First of all, we set the refractive index of the dielectric layers at the left and right edges to unity so that no (edge) reflections arise. As shown in Fig.~\ref{fig:gL}, the resistance, $g^{-1}$, increases linearly with the sample length, $L$. This confirms that the sample is a diffusive conductor. Indeed, the slope of the straight line in Fig.~\ref{fig:gL} is $6.33\times10^{-5}$, and noticing $g^{-1}=(L+2z_{b0})/\xi$ [cf. Eq.~(\ref{eq:52})] with $z_{b0}$ being the extrapolation length in the absence of edge reflection \cite{Rossum99,note1}, we obtain the localization length $\xi=1.58\times 10^4$ which is much larger than the sample length ($\xi/L=17.6$). Moreover, the intersection of the straight line with the vertical axis $L=0$ gives $2 z_{b0}/\xi = 9.52\times10^{-3}$. With the substitution of the value of $\xi$ we find $z_{b0}=75.31$.

\begin{table*}
\newcommand{\tabincell}[2]
{\begin{tabular}{@{}#1@{}}#2
\end{tabular}}
\centering
\caption{\label{Table2} The value of $z_b$ obtained by numerical simulations.}
\begin{tabular}{ccccccccccc}
  \hline\hline
  $\epsilon$\, & $1.0$ & $1.5$ & $1.6$ & $1.7$ & $1.9$ & $2.0$ & $2.1$ & $2.2$ & $2.3$ & $2.4$ \\
  \hline
  $z_b$($z_b'$)\, & $75.31$\, & $92.01$\, & $103.3$\, & $162.5$\, & $979.8$\, & $2111$\, & $4141$\, & $7504$\, & $12996$\, & $21123$\, \\
  \hline
  $\zeta$($\zeta'$)\, & $0.08$ & $0.10$ & $0.11$ & $0.18$ & $1.09$ & $2.35$ & $4.60$ & $8.34$ & $14.44$ & $23.47$ \\
  \hline\hline
\end{tabular}
\end{table*}

For fixed different values of refractive indexes at the left and right edges, by varying the sample length and computing $g$ we also confirm the general expression (\ref{eq:52}). Importantly, this result allows us to determine the value of the length $z_b$ for different values of the dielectric constant $\epsilon$ of edge dielectric layer. Specifically, we fix $\epsilon$ of the dielectric layer at the left edge to be unity while vary that at the right. For each dielectric
constant value corresponding to the right edge we compute the average conductance for different sample lengths. As mentioned above, $g^{-1}$ increases linearly with $L$. For this straight line, we find that, as shown in Fig.~\ref{fig:gL}, the slope is independent of dielectric
constant at the right edge, in agreement with the prediction of Eq.~(\ref{eq:52}), whereas the intersection with the vertical axis $L=0$ varies. Subtracting $z_{b0}$ from the product of $\xi$ and this intercept, we find $z_b$ for corresponding $\epsilon$ at the right edge. The results of $z_b$ for different values of $\epsilon$ are given in Table \ref{Table2}.

\subsection{DTE transition}
\label{sec:transition}

\begin{figure}
 \begin{center}
 \includegraphics[width=8.0cm]{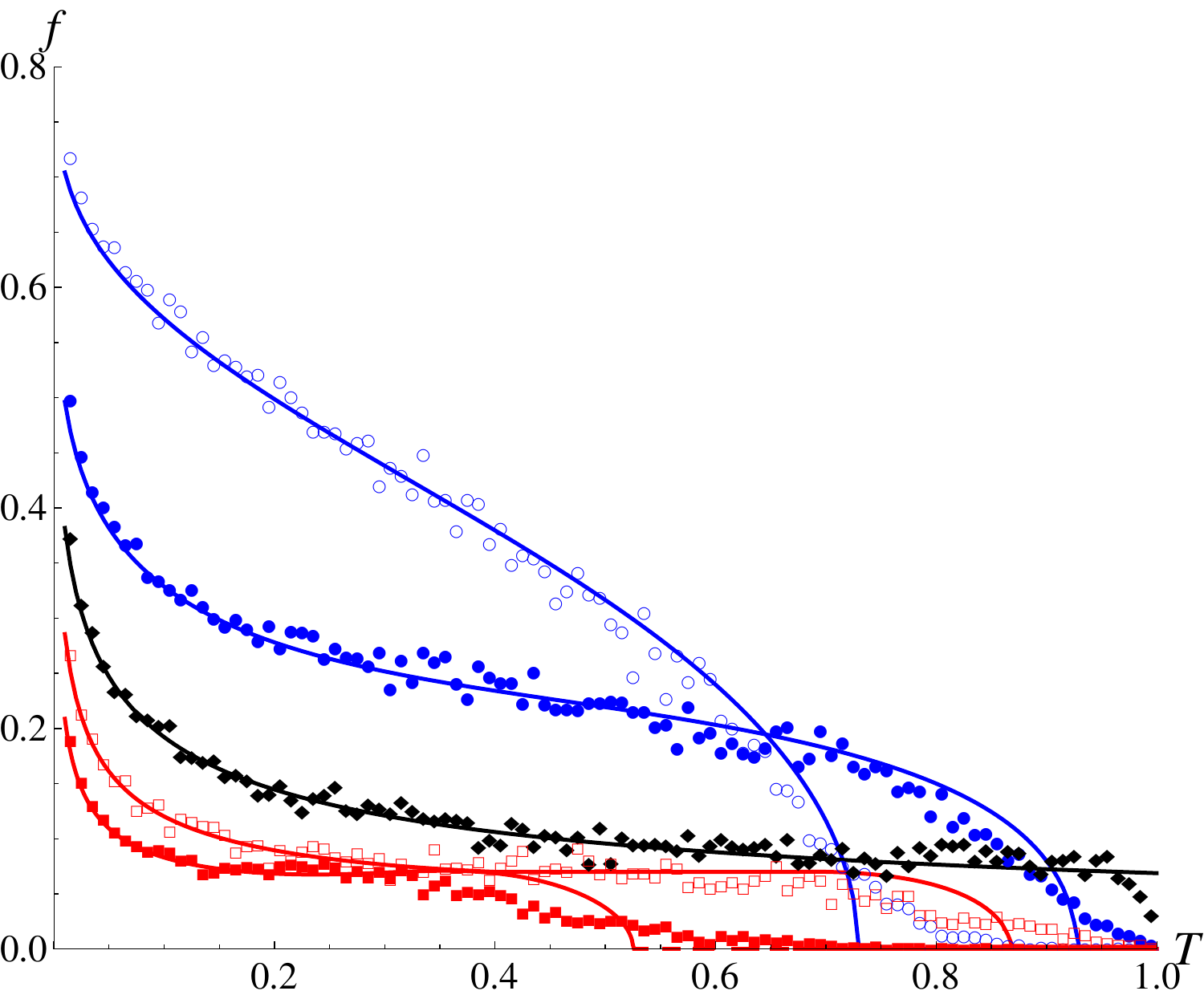}
\end{center}
 \caption{Simulations confirm the occurrence of double phase transitions. The dielectric constant of the reflection layer at the left edge is fixed to $\epsilon'=2.1$ and that at right, $\epsilon$, increases. The simulation results are for $\epsilon=1.0$ (blue empty circles), $1.9$ (blue solid circles), $2.1$ (black solid diamonds), $2.3$ (red empty squares) and $2.4$ (red solid squares), respectively. The blue (red) circles correspond to the upper left (lower right) $C$-phase regime in Fig.~\ref{fig:8} where the high transmission eigenchannels are blocked; the black circles correspond to the $O$-phase where the high transmission eigenchannels are present. The simulation results are in good agreement with the analytical results of $f({\cal T})$ (solid lines) without any fitting parameters.}
 \label{fig:4}
\end{figure}

We now focus on the most interesting case of double phase transitions.
To numerically explore the DTE transition we fix the dielectric constant of the reflection layer at the
left edge to $2.1$ (corresponding to $\zeta'=4.60$, see Table \ref{Table2}) and tune that at the right edge from $1$ to $2.4$ with the increment of $0.1$. Because of $\zeta'>1$ the corresponding phase structure must be the same as Line II in Fig.~\ref{fig:8}.
According to the exact criterion (\ref{eq:21}) and the numerical value of $\zeta$ given in Table \ref{Table2} we predict that the system undergoes double phase transitions when $\epsilon$ increases and takes the values in this table: for $\epsilon$ less than (or equal to) $2.0$, because of $\zeta'-\zeta>1$ the system is in the {\it C}-phase (upper left regime in Fig.~\ref{fig:8}); for $\epsilon=2.1$, because of $\zeta'-\zeta=0$ the system is in the {\it O}-phase;
for $\epsilon$ larger than (or equal to) $2.2$, because of $\zeta-\zeta'>1$ the system enters into the other {\it C}-phase regime (lower right regime in Fig.~\ref{fig:8}). These double phase transitions are observed numerically, as shown in Fig.~\ref{fig:4}, where the colors of blue, red, and black correspond to two {\it C}-phases in different regimes of Fig.~\ref{fig:8} and the {\it O}-phase, respectively. (To find the critical phase is beyond our current numerical experimental reach since this would require the fine tuning of $\epsilon$ at the sample edge.)

Moreover, the analytic results of the $f$-factor (solid lines), with the values of $z_b,z_b'$ given by Table \ref{Table2}, are in good agreement with simulation results without any fitting parameters. The smearing of the singularity is an effect of finite number of channels. In simulations, the transmission eigenvalue density at ${\cal T}$ is defined as the ratio of the number of eigenvalues in the interval of $[{\cal T}-\frac{1}{2}\Delta{\cal T},{\cal T}+\frac{1}{2}\Delta{\cal T}]$ ($\Delta{\cal T}$ set to $0.01$) to the total number of eigenvalues.
Therefore, numerical experiments confirm that for
sufficiently large $\epsilon'$, so that $\zeta'>1$, the system undergoes double DTE transitions as $\epsilon$ at the right edge increases from unity.

As we have analytically shown, the phase structure corresponding to Line I in Fig.~\ref{fig:8}, where the single phase transition occurs, is similar to its limiting case corresponding to the line of $\zeta'=0$ in Fig.~\ref{fig:8}. The latter has been confirmed in numerical simulations previously \cite{Tian13}. So we do not discuss single phase transition here.

\section{Physical mechanism}
\label{sec:physical_interpretation}

Above we have developed an exact analytical theory for the DTE transition and confirmed the transition by numerical simulations. In this section we further provide a transparent physical picture based on combined analytical and numerical analyses.

\subsection{Mapping to resonator model}
\label{sec:resonant_cavity_model}

We first note that, very recently, it has been discovered \cite{Genack15} that the structure of eigenchannels, namely, the (spatial) profile of the wave energy density, is universal. It is governed by the localization length associated with the corresponding transmission eigenvalue \cite{note_Dorokhov} and the (position-dependent) diffusion coefficient \cite{Tian10} (see also Ref.~\onlinecite{Tian13a} for a review).
Based on this fact it is natural to expect the eigenvalue, i.e., the parameter $\phi$ in Eq.~(\ref{eq:33}) depends implicitly on a certain position variable, $x$, associated with the eigenchannel structure.
Loosely speaking, this variable characterizes how deep the energy is pumped into the medium. Its physical meaning will become clearer below.

\subsubsection{$\phi$-resonator center correspondence}
\label{sec:center}

Let us hypothesize (see Appendix~\ref{sec:parametrization} for more discussions) that the explicit dependence of $\phi(x)$ is
\begin{equation}\label{eq:34}
    \phi\equiv 2(L-2x)/\ell.
\end{equation}
Recall that the left edge of the sample corresponds to $x=0$. Then,
\begin{equation}\label{eq:64}
    {\cal T}=\cosh^{-2}\frac{L-2x}{\ell}
\end{equation}
and
\begin{equation}\label{eq:35}
    \frac{\ell}{4L}\frac{d{\cal T}}{{\cal T}\sqrt{1-{\cal T}}}=d\left(\frac{x}{L}\right).
\end{equation}
This shows that a bimodal distribution of ${\cal T}$ is equivalent to a uniform distribution of the position variable $x$. Indeed, the distribution given by the left-hand side of Eq.~(\ref{eq:35}) differs from Eq.~(\ref{eq:20}) only in an unimportant overall factor. More precisely, it can be shown that provided $x$ is uniformly distributed over the sample, then the mapping: $x\rightarrow \phi$ leading to the bimodal distribution is {\it unique} (see Appendix~\ref{sec:parametrization} for the proof), and is given by Eq.~(\ref{eq:34}).

According to Eqs.~(\ref{eq:34}) and (\ref{eq:64}), the eigenvalue ${\cal T}=1$ corresponds to $x=L/2$. In the presence of edge reflection, Eq.~(\ref{eq:52}) suggests that effectively the sample length is extended by an amount of $z_b'$ from the left edge and of $z_b$ from the right. Taking this taken into account, we modify Eq.~(\ref{eq:64}) as
\begin{equation}\label{eq:36}
    {\cal T}=\cosh^{-2}\frac{(L+z_b'+z_b)-2(x+z_b')}{2\ell},
\end{equation}
and the eigenvalue ${\cal T}=1$ corresponds to
\begin{equation}\label{eq:50}
    x=(L+z_b-z_b')/2.
\end{equation}
Because $0\leq x\leq L$, we find
\begin{equation}\label{eq:15}
   |z_b-z_b'|\leq L.
\end{equation}
This is identical to the inequality (\ref{eq:12}) for which perfectly transmitting eigenchannels are present.
Therefore, the criterion for the DTE transition (\ref{eq:21}) is reproduced.

\begin{figure}
 \begin{center}
 \includegraphics[width=8.0cm]{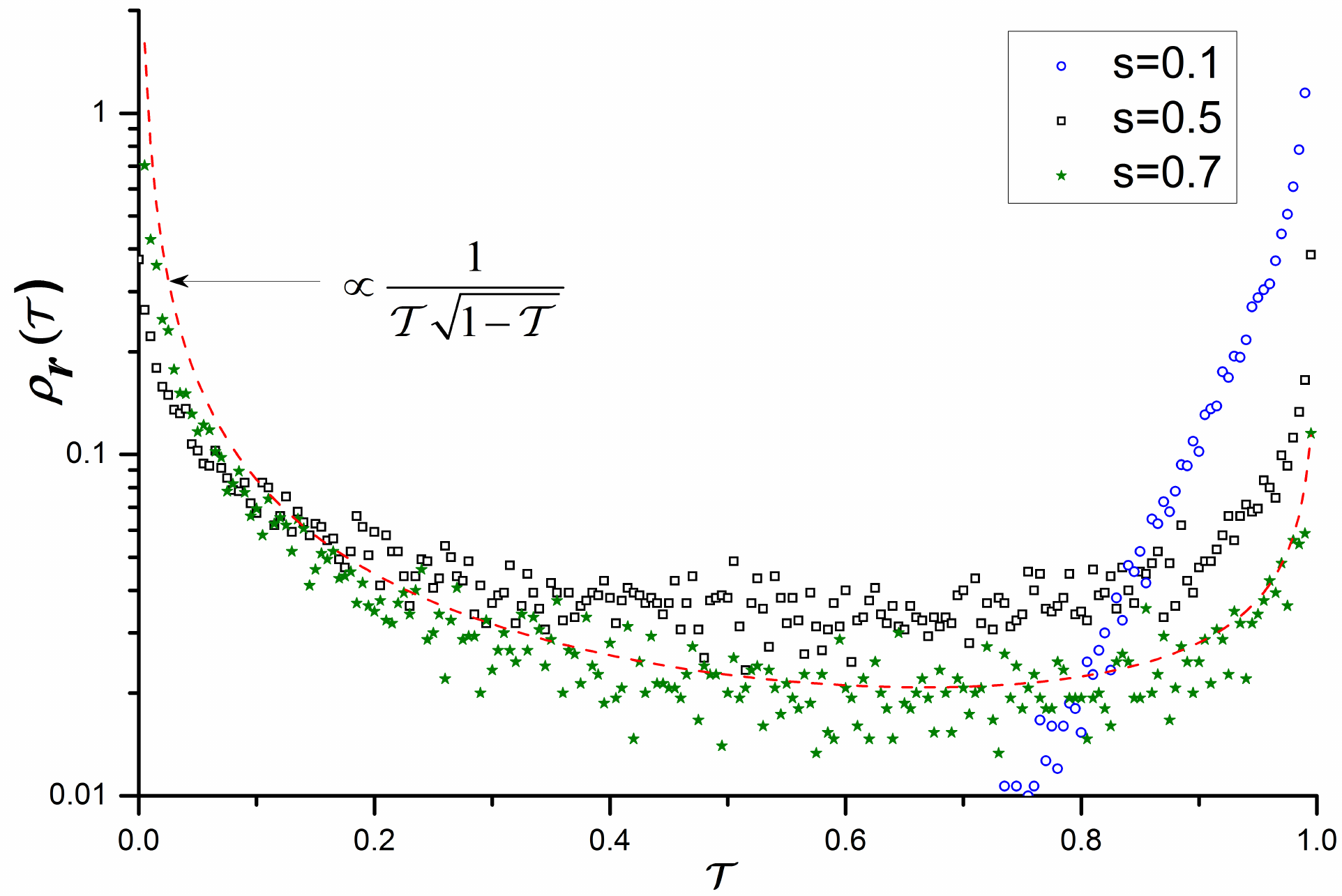}
\end{center}
 \caption{Simulations show that for sufficiently strong disorders (large $s$) DRT exhibits bimodal behavior. The samples are $1$D with transparent edges.}
 \label{fig:5}
\end{figure}

To better understand the physical meaning of the position variable $x$ we note that the parametrization (\ref{eq:64}) has an equivalent expression,
\begin{equation}\label{eq:13}
    {\cal T}(x)=\frac{4T'(x)T(x)}{[T'(x)+T(x)]^2},
\end{equation}
with
\begin{equation}\label{eq:14}
    T'(x)=
    e^{-2x/\ell},\quad T(x)=
    e^{-2(L-x)/\ell}.
\end{equation}
Surprisingly, Eq.~(\ref{eq:13}) is identical to the expression for the transmittance through a disorder-induced $1$D resonator in $1$D random media in the localized regime, where the localization length is order of the transport mean free path \cite{Freilikher03}. [In fact, the expression (\ref{eq:13}) is similar to the general expression (\ref{eq:44}) below for resonant transmittance.] Because $x,L-x$ are much larger than $\ell$, the resonator corresponding to ${\cal T}(x)$ given by Eq.~(\ref{eq:13}) essentially is a point-like defect, placed at $x$ and confined within a size $\sim {\cal O}(\ell)$. This analogy between the transmission eigenvalue and resonant transmittance suggests that the (high) transmission eigenchannels have a close relation to
resonators and therefore are of strong interference origin. The mapping of a high transmission eigenchannel in the system of higher dimension onto $1$D resonator makes sense perhaps because the eigenvalue ${\cal T}$ corresponds to the universal eigenchannel profile, and this profile is structureless in the transverse direction of the waveguide \cite{Genack15}.

\subsubsection{Analog of $f$-factor in resonator model}
\label{sec:f}

As shown in Eq.~(\ref{eq:35}), in samples without edge reflections, the bimodal distribution follows from a uniform distribution of resonator centers. To find out what happens in systems with reflective edges, we multiple both sides of Eq.~(\ref{eq:35}) by the deviation factor $f$ to obtain
\begin{equation}\label{eq:37}
    \frac{\ell}{2L}\frac{f({\cal T})d{\cal T}}{{\cal T}\sqrt{1-{\cal T}}}=f[{\cal T}(x)]d\left(\frac{x}{L}\right).
\end{equation}
The left-hand side gives the general expression of DTE, i.e., $\rho({\cal T})=f({\cal T})\rho_0({\cal T})$; the right-hand side shows that the physical meaning of the $f$-factor is the spatial density of resonators. It then becomes clear that the closing of the perfectly transmitting eigenchannel can be interpreted as the vanishing of resonators with $T'(x)=T(x)$. Indeed, if the reflections of the left and right edges are so asymmetric that an effective resonator with $T'(x)=T(x)$ cannot be introduced, then according to Eq.~(\ref{eq:13}) the maximum transmission eigenvalue must be smaller than unity.

\subsubsection{Normalization condition in resonator model}
\label{sec:normalization}

We note that the bimodal distribution is not normalizable. Because the integral $\int_0^1 \rho_0({\cal T})d{\cal T}$ suffers logarithmic divergence due to the ${\cal T}^{-1}$ singularity for ${\cal T}\rightarrow 0$. Therefore, we cut the integral at a certain exponentially small value, $T_c\approx 4e^{-2L/\ell}$, so that $\int_{T_c}^1 \rho_0({\cal T})d{\cal T}=N$. On the other hand, from Eq.~(\ref{eq:13}) it is easy to see that
\begin{equation}\label{eq:38}
    {\cal T}(x) \gtrsim T_{typ} = 4e^{-2L/\ell}.
\end{equation}
This implies that the cutoff of the bimodal distribution, $T_c$, has the meaning of the typical transmission, $T_{typ}$, in the resonator model.

\begin{figure}
 \begin{center}
 \includegraphics[width=8.0cm]{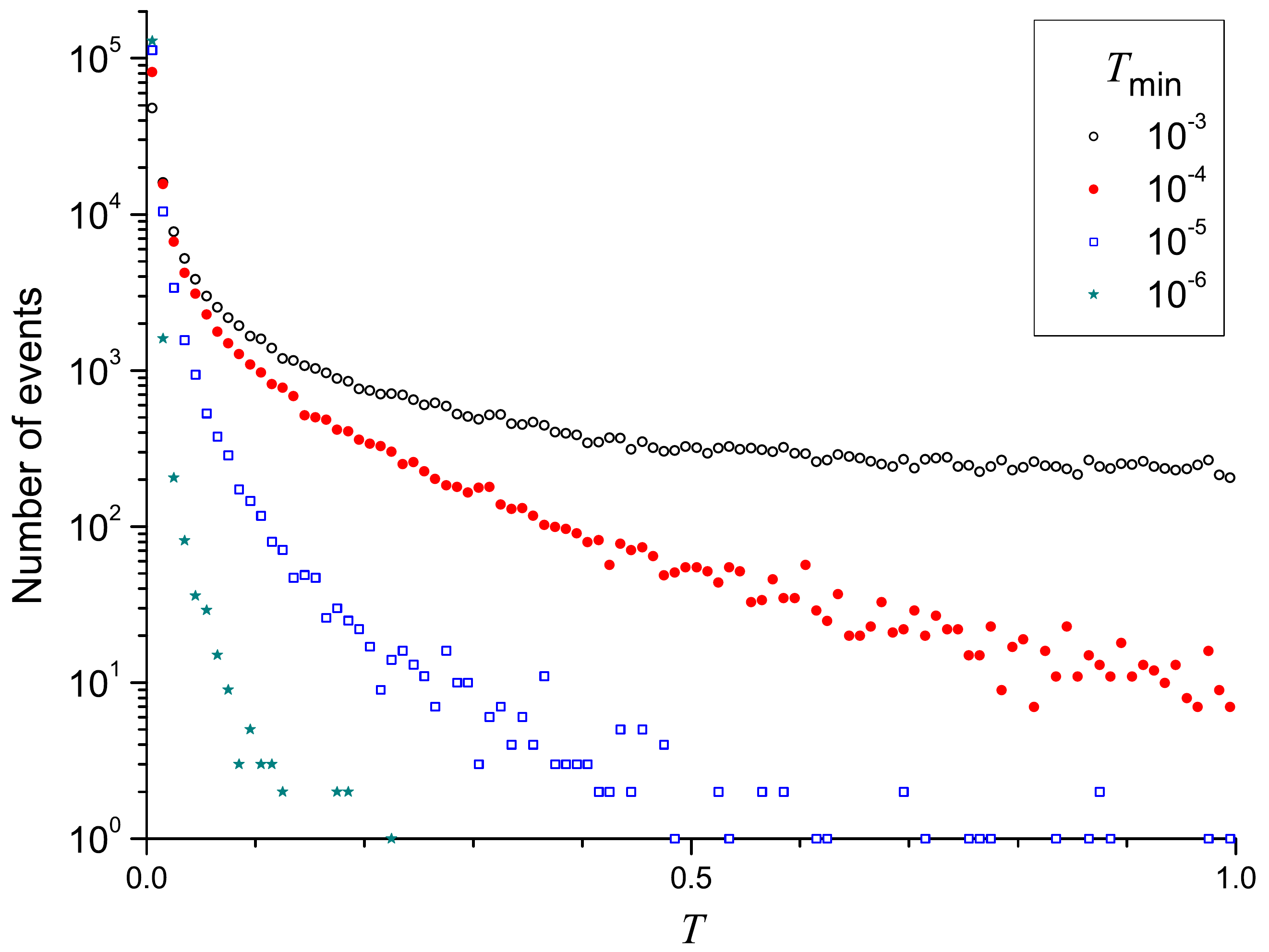}
\end{center}
 \caption{Simulations of $1$D samples show that the perfect resonant transmittance
 (${\cal T}=1$) disappears when the reflection coefficient
 ($1-T_{\rm min}$) of one sample edge (the other is always transparent)
 exceeds a critical value. The disorder strength parameter $s=0.5$. The vertical axis is the number of events and the distribution, therefore, is not normalized.
 }
 \label{fig:6}
\end{figure}

\subsection{Transition in distribution of the resonant transmittance (DRT)}
\label{sec:resonant_cavity_model_numerical}

From analysis above we have seen that although the transmission eigenvalue and resonant transmittance are very different concepts, they have well-pronounced similarity. Naturally, we expect the DRT to exhibit a transition similar to that of DTE. In this part we will address this issue. We perform numerical experiments on wave transmission through $1$D random media, namely layered samples in the strong localization regime.

\subsubsection{Numerical observations of DRT transition}
\label{sec:DRT_numerical_observation}

We randomly place a fixed amount (set to $50$ in numerical experiments below) of scatterers in a $1$D chain, and let the reflection coefficients of scatterers be $sr_i$ ($i$ labels the scatterers).
For each $i$, $r_i$ randomly takes a value from the interval of $(-1,1)$. The constant $s$ satisfying $0<s<1$ is independent of $i$.
The disorder strength is controlled by the value of $s$. The dimensionless distances 
between nearest scatters are randomly distributed in the interval $\delta_0\pm\Delta\delta$,  $\Delta \delta/\delta_0=90\%$. We change the frequency $\omega$ of incident waves in the narrow interval $\omega_0\pm\Delta\omega$, $\Delta\omega/\omega_0=5\%$, and calculate the transmittance spectrum
${\cal T}(\omega)$ by using the standard transfer matrix approach. The numerical experiments are repeated for $10^3$ disorder configurations. There are about $10^2$ resonant peaks (transmittance resonances) in the spectrum ${\cal T}(\omega)$ in any configurations in the given frequency interval, so that the total number of resonances analyzed below is about $10^5$.


First, we study the case where both sample edges are transparent. We find that the background value of the spectrum
${\cal T}(\omega)$, namely the typical transmission coefficient, is very close to zero in accord with the expression of $T_{typ}$ given in Eq.~(\ref{eq:38}). The resonator corresponds to the local maximum in this transmittance spectrum, i.e., ${\cal T} = {\rm max}\{{\cal T}(\omega)\}$. As shown in Fig.~\ref{fig:5}, for sufficiently strong disorder strengths (so that the resonator centers are uniformly distributed in the sample) the DRT, denoted as $\rho_r({\cal T})$, is in good agreement with the distribution (\ref{eq:20}) while for moderate disorder strengths $\rho_r({\cal T})$ deviates from the behavior of Eq.~(\ref{eq:20}) but is still bimodal.

Next, we study effects of edge reflection on the DRT. For simplicity we consider only the case where one (say the left) edge is transparent while the reflection of the other edge, denoted as $(1-T_{\rm min})$, increases. The simulation results are shown in Fig.~\ref{fig:6}. We see that when the reflection increases and exceeds certain critical value, the peak at ${\cal T}=1$ is fully suppressed, i.e., the perfect resonant transmittance (${\cal T}=1$) disappears. When the internal reflection further increases, the highest resonant transmittance decreases. This behavior is the same as the single transition exhibited by the DTE for quasi $1$D disordered waveguides with one edge transparent and the other reflective.

\begin{figure}
 \begin{center}
 \includegraphics[width=8.0cm]{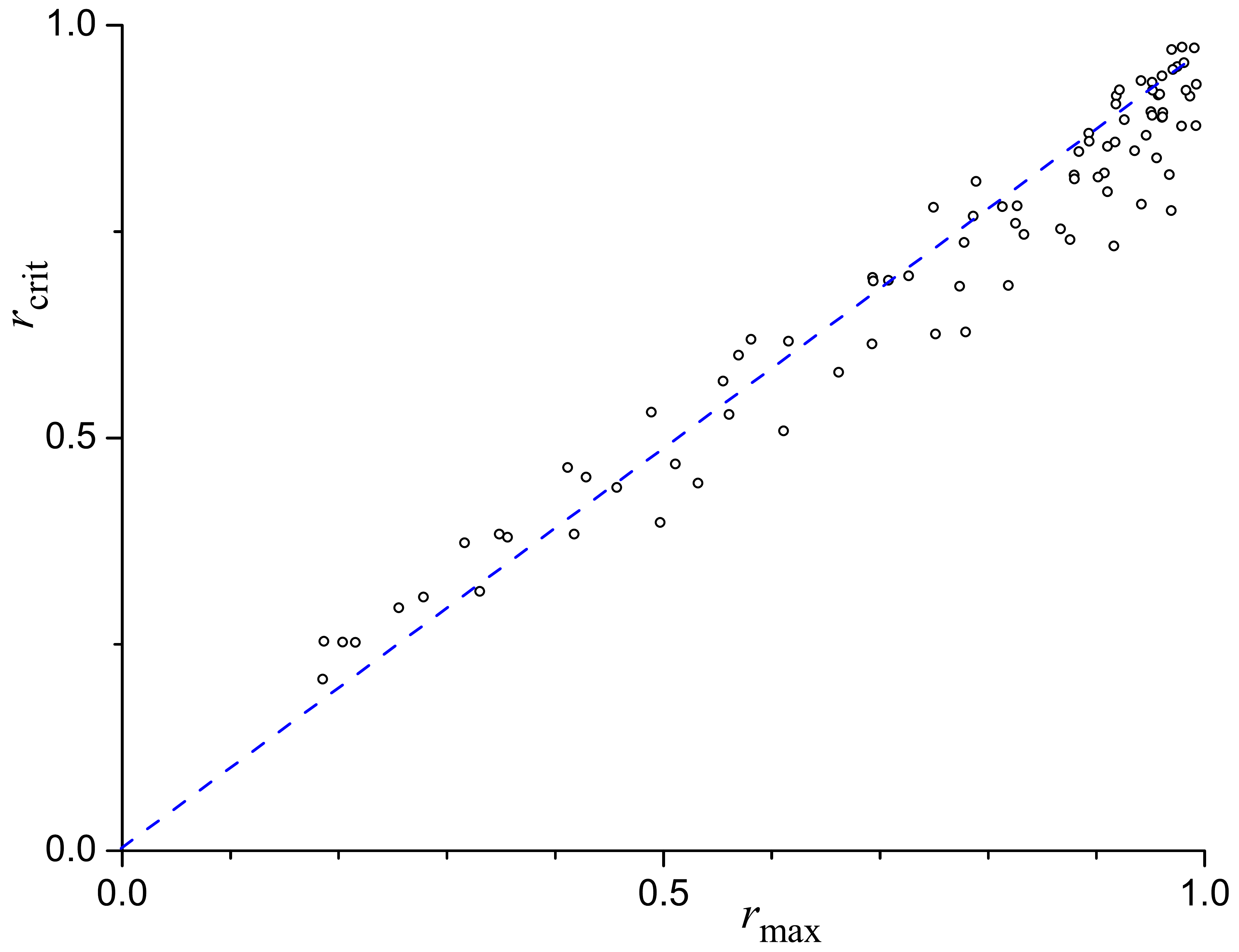}
\end{center}
 \caption{Simulations confirm that the critical value of the edge reflector $r_{\rm crit}$, above which perfect resonant transmittances disappear, is the same as the maximal value $r_{\rm max}$ of $|r_s(\omega)|$ for the same sample without edge reflector.}
 \label{fig:7}
\end{figure}

\subsubsection{Physical mechanism for DRT transition}
\label{sec:DRT_origin}

We show below that the DRT transition can be well explained by the resonant cavity theory for $1$D disordered systems \cite{Freilikher03}. Suppose that the $1$D sample without edge reflection has a complex reflection coefficient $r_s(\omega)$. This sample itself may be considered as a single semitransparent barrier, characterized by $r_s(\omega)$. Upon placing additional reflector after the output edge of the sample, which is described by a complex $\omega$-independent reflection coefficient $r_{\rm end}$, we have two barriers that constitute a resonator. The total transmittance of this resonator is
\begin{eqnarray}
    {\cal T}(\omega)&=&\frac{4T_s(\omega)T_{\rm min}}{\left(T_s(\omega)+T_{\rm min}\right)^2},\label{eq:44}\\
    T_s(\omega)&=&1-|r_s(\omega)|^2,\quad T_{\rm min}=1-|r_{\rm end}|^2.
\end{eqnarray}
This shows that the resonator can be perfectly transparent only if $|r_s(\omega)|=|r_{\rm end}|$ and the round trip phase shift is multiple of $2\pi$. When $|r_{\rm end}|$
is small enough and the frequency varied in a rather broad range, there are many frequencies for
which these conditions are satisfied. How does the number of such frequencies, $N_c$, evolve with increasing $|r_{\rm end}|$? Obviously, the number of frequency regions where $|r_s(\omega)|=|r_{\rm end}|$ becomes smaller and $N_c$ decreases. When $|r_{\rm end}|$ exceeds maximal value $r_{\rm max}$ of $|r_s(\omega)|$ (for given sample in given frequency range), the condition $|r_s(\omega)|=|r_{\rm end}|$ cannot be satisfied by any frequency. Thus, $r_{\rm max}$ is the critical value of the edge reflector above which perfect resonant transmittances disappear.

To check this result numerically, we firs find $r_{\rm max}$ for all $1$D random samples. Then, we add additional layer with reflection coefficient $r_{\rm end}$ and select those ${\cal T}(\omega)$ larger than $0.999$ as perfect resonant transmittances. Then, we increase $|r_{\rm end}|$
until no frequencies in the given range satisfy this selection rule. This value, denoted as $r_{\rm crit}$, is the critical value for the DRT transition. As shown in Fig.~\ref{fig:7}, simulations confirm $r_{\rm crit}=r_{\rm max}$.

\subsection{Physical picture of DTE transition}
\label{sec:physical_picture_DTE}

Based on the demonstrated similarities between the transmission eigenvalues and resonant transmittances, and  between the DTE and DRT transitions, in what follows, we propose a physical explanation of the DTE transition.

As shown in Ref.~\onlinecite{Genack15}, for quasi $1$D samples without edge reflections, the eigenchannel structure corresponding to perfect transmission is given by
\begin{equation}\label{eq:48}
    W_{{\cal T}=1}(x)=1+F_1(x),
\end{equation}
where $F_1(x)$ is essentially the probability density for a wave to return to a cross section at depth $x$ in an open random medium. (We put the left edge as the input edge.) The independence of the eigenchannel profile $W_{\cal T}(x)$ on the transverse coordinate reflects the $1$D nature of the universal eigenchannel structure. For diffusive samples without edge reflections, $F_1(x)$ has an explicit form of $\pi x(L-x)/(2L\ell)$ which is symmetric with respect to the sample center (Fig.~\ref{fig:9}, red curve in upper panel). This quadratic form of $F_1(x)$ can be obtained by solving the (normal) diffusion equation with the Dirichlet boundary condition \cite{Genack15}. In the presence of edge reflections, the diffusion equation is the same but the boundary conditions are of the mixed type. Such diffusion equation can be easily solved, giving
$F_1(x)=\pi (x+z_b')(L-x+z_b)/(2(L+z_b'+z_b)\ell)$. Upon substituting this expression into Eq.~(\ref{eq:48}) we find (Fig.~\ref{fig:9}, blue curve in upper panel)
\begin{equation}\label{eq:49}
    W_{{\cal T}=1}(x)=1+\frac{\pi (x+z_b')(L-x+z_b)}{2(L+z_b'+z_b)\ell}.
\end{equation}
When $z_b,z_b'$ vanish Eq.~(\ref{eq:49}) reduces to the result obtained in Ref.~\onlinecite{Genack15}. In Appendix~\ref{sec:diffuson} we further discuss a relation between this eigenchannel profile and the field theory (\ref{eq:3}).

The position of the center of the profile (\ref{eq:49}) is precisely the same as what is given by Eq.~(\ref{eq:50}).
That is, for this eigenchannel (${\cal T}=1$) the resonator center is locked to the profile center. So, if the reflection of the right (left) edge is stronger than that at the left (right), i.e., $z_b>z_b'$ ($z_b<z_b'$) the profile
and resonator centers move to the right (left) off
the sample center (cf. Fig.~\ref{fig:9}). As the
asymmetry parameter continues increasing,
eventually the resonator center approaches the sample edge,
and the perfectly transmitting eigenchannel disappears, signaling the DTE transition.
We emphasize that although the resonator is
due to the interference of multiply scattered random fields, it exists in the diffusive sample, far from the localization regime.

\section{Conclusion and discussion}
\label{sec:conclusions}

We have shown analytically and confirmed numerically that asymmetry in the edge reflections has a significant impact on the wave propagation through disordered media. In particular, it can trigger a peculiar DTE double-transition phenomenon in quasi $1$D diffusive samples. When the asymmetry in the reflections of the two sample edges is weak so that the inequality $|\zeta-\zeta'|<1$ is satisfied (i.e, the difference between the two edge resistances is smaller than the bulk resistance), high transmission eigenchannels are present, and the DTE exhibits a singularity $\rho({\cal T}\rightarrow 1)\sim (1-{\cal T})^{-\frac{1}{2}}$. In the opposite case of strong asymmetry (i.e, the difference between the two edge resistances is larger than the bulk resistance), high transmission eigenchannels are closed. When the asymmetry parameter $|\zeta-\zeta'|=1$ (i.e, the difference between the two edge resistances equals to the bulk resistance), the system is in a critical phase with the DTE transition exhibiting critical statistics, i.e., $\rho({\cal T}\rightarrow 1)\sim (1-{\cal T})^{-\frac{1}{3}}$. These phenomena are universal in the sense that they are governed by a single parameter $|\zeta-\zeta'|$ and at each phase regime, the asymptotic behavior of $\rho({\cal T}\rightarrow 1)$ does not depend on the details of system's structure such as the specific value of edge reflection, disorder configuration, sample width, etc..

Surprisingly, notwithstanding its occurrence in diffusive samples, far from the localization regime, the DTE transition has a close connection to the resonator model of strong localization in $1$D. More precisely, the DTE can be mapped onto the statistics of the resonator center so that the factor $f(=\rho/\rho_0)$ corresponds to the spatial density of resonators. We show that perfectly transmitting eigenchannels can be modeled by $1$D resonators, and the disappearance of such an eigenchannel as the asymmetry in the edge reflections increases mimics the evolution of a transmittance resonance when the corresponding effective cavity shifts to an edge of the sample.

Our findings indicate a novel coherent phenomenon of diffusive waves. They exist in very general wave systems, e.g., quantum matter and classical elastic waves, wherever edge reflection is strong. A prominent system is the normal-metal -- superconductor junction \cite{Beenakker97}, where edge reflection is created by the tunnel barrier at the normal-metal -- superconductor interface. Another system is the dissimilar solid \cite{Little59} where the Kapitza resistance on
the thermal phonon transport are not negligible. Our findings may enable the control of transmission eigenchannels and eigenvalues via edge reflection, which is well within experimental reach. The indication of the existence of resonators in diffusive samples may find practical applications such as low-threshold lasing.

An important subject of further studies in this
direction is to better understand the
formation of resonators in quasi $1$D diffusive random media. A related issue is the connection
between the resonator and the eigenchannel
structure for arbitrary transmission eigenvalue. This would help to uncover the physical meaning of the linear connection (\ref{eq:34}) between the parameter $\phi$ and
 the resonator center $x$. The subject of particular interests is the interplay between edge reflection and gain (or absorption).

Our investigations of the DTE transition have been restricted to quasi $1$D diffusive samples. We should emphasize that the microscopic formalism, i.e., the supersymmetry field theory of DTE presented in Sec.~\ref{sec:transition_diffusive}, can be directly applied to localized samples. The additional technical difficulty in calculating Eq.~(\ref{eq:3}) is that one needs to take into account all the saddle point configurations in which the supersymmetry is broken. How would the DTE transition be affected then? We leave this challenging problem for future studies.

\section*{Acknowledgements}

We would like to thank X. J. Cheng, M. Davy, A. Z Genack, and Z. Shi for useful discussions. This work is supported by the NSFC (No. 11174174) and
by the Tsinghua University ISRP.

\begin{appendix}

\section{Derivations of Eqs.~(\ref{eq:7}) and (\ref{eq:8})}
\label{sec:calculations}

We rewrite Eq.~(\ref{eq:4a}) as
\begin{widetext}
\begin{equation}
\label{eq:4c}
\begin{split}
2 \overline{C}_\phi^2 +\Delta C_\phi^2/2
&=\dfrac{\sinh^2 \psi_+ (a \cosh\psi_- +b) + \sinh^2 \psi_- (a \cosh\psi_+ +b)}{(a \cosh\psi_+ +b)(a \cosh\psi_- +b)}\\
&=\dfrac{a(\sinh^2\psi_+ \cosh\psi_- +\sinh^2\psi_- \cosh\psi_+)+b(\sinh^2\psi_+ +\sinh^2\psi_-)}
{a^2 \cosh\psi_+ \cosh\psi_- +ab(\cosh\psi_+ +\cosh\psi_-) + b^2},
\end{split}
\end{equation}
and Eq.~(\ref{eq:4b}) as
\begin{equation}
\label{eq:4e}
\begin{split}
2 \overline{C}_\phi \Delta C_\phi
&=\dfrac{\sinh^2 \psi_+ (a \cosh\psi_- +b) - \sinh^2 \psi_- (a \cosh\psi_+ +b)}{(a \cosh\psi_+ +b)(a \cosh\psi_- +b)}\\
&=\dfrac{a(\sinh^2\psi_+ \cosh\psi_- -\sinh^2\psi_- \cosh\psi_+)+b(\sinh^2\psi_+ -\sinh^2\psi_-)}
{a^2 \cosh\psi_+ \cosh\psi_- +ab(\cosh\psi_+ +\cosh\psi_-) + b^2}.
\end{split}
\end{equation}
\end{widetext}
To proceed further we use the following identities:
\begin{align}
\cosh\psi_+ \cosh\psi_- &=
(\cosh2\bar{\psi} +\cosh\Delta\psi)/2,\nonumber\\
\cosh\psi_+ +\cosh\psi_- &=
2\cosh\bar{\psi}\cosh\dfrac{\Delta\psi}{2},\nonumber\\
\cosh\psi_+ -\cosh\psi_- &=
2\sinh\bar{\psi}\sinh\dfrac{\Delta\psi}{2},\nonumber
\end{align}
with $
\bar \psi=(\psi_++\psi_-)/2=\overline{C}_\phi-\phi,\quad\Delta \psi=\psi_+-\psi_-=\Delta C_\phi-2i\pi$, to obtain
\begin{eqnarray}
\label{eq:4d}
&&\sinh^2\psi_+ \cosh\psi_- +\sinh^2\psi_- \cosh\psi_+ \nonumber\\
&=&(\cosh2\bar{\psi} +\cosh\Delta\psi-2)\cosh\bar{\psi}\cosh\dfrac{\Delta\psi}{2}
\end{eqnarray}
\begin{eqnarray}
\label{eq:4g}
&&\sinh^2\psi_+ \cosh\psi_- -\sinh^2\psi_- \cosh\psi_+ \nonumber\\
&=& (\cosh2\bar{\psi} +\cosh\Delta\psi+2)\sinh\bar{\psi}\sinh\dfrac{\Delta\psi}{2},
\end{eqnarray}
\begin{eqnarray}
\label{eq:4f}
\sinh^2\psi_+ +\sinh^2\psi_-
=\cosh2\bar{\psi}\cosh\Delta\psi-1,
\end{eqnarray}
and
\begin{eqnarray}
\label{eq:4h}
\sinh^2\psi_+ -\sinh^2\psi_-
= \sinh2\bar{\psi}\sinh\Delta\psi.
\end{eqnarray}
Substituting them into Eqs.~(\ref{eq:4c}) and (\ref{eq:4e}) gives
\begin{widetext}
\begin{equation}\label{eq:6}
2 \overline{C}_\phi^2 +\Delta C_\phi^2/2 =\dfrac{a(\cosh2\bar{\psi} +\cosh\Delta\psi-2)\cosh\bar{\psi}\cosh\dfrac{\Delta\psi}{2}+b(\cosh2\bar{\psi}\cosh\Delta\psi-1)}
{a^2(\cosh2\bar{\psi}+\cosh\Delta\psi)/2+2ab\cosh\bar{\psi}\cosh\dfrac{\Delta\psi}{2}+b^2},
\end{equation}
\begin{equation}\label{eq:5}
2 \overline{C}_\phi \Delta C_\phi =\dfrac{a(\cosh2\bar{\psi} +\cosh\Delta\psi+2)\sinh\bar{\psi}\sinh\dfrac{\Delta\psi}{2}+b\sinh2\bar{\psi}\sinh\Delta\psi}
{a^2(\cosh2\bar{\psi}+\cosh\Delta\psi)/2+2ab\cosh\bar{\psi}\cosh\dfrac{\Delta\psi}{2}+b^2},
\end{equation}
which are Eqs.~(\ref{eq:7}) and (\ref{eq:8}) upon the substitutions of $\Delta C_\phi$, $\bar{\psi}$, and $\Delta\psi$.

\section{Values of $\alpha$ and $\beta$}
\label{sec:expansion}

Taking into account $b-a = 1$ and introducing $\overline{f} \equiv \pi f$, we rewrite Eqs.~(\ref{eq:7}) and (\ref{eq:8}) as
\begin{equation}
\label{eq:5a}
2 ({\bar \psi}^2 -\overline{f}^2) = \dfrac{-a(\cosh{2{\bar \psi}} +\cos{2\overline{f}}-2)\cosh{{\bar \psi}}\cos{\overline{f}}+(a+1)(\cosh{2{\bar \psi}}\cos{2\overline{f}}-1)}
{a^2(\cosh{2{\bar \psi}}+\cos{2\overline{f}})/2-2a(a+1)\cosh{{\bar \psi}}\cos{\overline{f}}+(a+1)^2},
\end{equation}
and
\begin{equation}
\label{eq:5b}
4({\bar \psi} +\phi) \overline{f} =\dfrac{-a(\cosh{2{\bar \psi}} +\cos{2\overline{f}}+2)\sinh{{\bar \psi}}\sin{\overline{f}}+(a+1)\sinh{2{\bar \psi}}\sin{2\overline{f}}}
{a^2(\cosh{2{\bar \psi}}+\cos{2\overline{f}})/2-2a(a+1)\cosh{{\bar \psi}}\cos{\overline{f}}+(a+1)^2}.
\end{equation}
\end{widetext}
As shown in Sec.~\ref{sec:critical_scaling}, the numerical solution to Eqs.~(\ref{eq:7}) and (\ref{eq:8}) yields the general form,
\begin{equation}\label{eq:27}
    \overline{f} =k_1 \phi^\alpha, \quad {\bar \psi} =k_2 \phi^\beta, \quad \phi \rightarrow 0,
\end{equation}
where $k_{1,2}$ are coefficients depending on $a$. The numerical solution further yields $\alpha\approx \beta\approx 1/3$.

We now show that Eq.~(\ref{eq:27}) and the relation above for $\alpha$ and $\beta$ lead to a stronger result. Let us expand Eq.~(\ref{eq:5a}) around $\phi=0$. Because of Eq.~(\ref{eq:27}), the first few order terms of $\phi$ in Eq.~(\ref{eq:5a}) are $\phi {\bar \psi}$, $\overline{f}^4$, ${\bar \psi}^2\overline{f}^2$ and
$\overline{f}^4$. At this stage we cannot determine for these terms which one is smaller, but all the other terms in the expansion must be of higher order because of $\alpha\approx \beta\approx 1/3$. We perform the same expansion for Eq.~(\ref{eq:5b}), for which the first few order terms are $\phi \overline{f}$, $\overline{f}^3{\bar \psi}$ and
$\overline{f}{\bar \psi}^3$.
As a result,
\begin{eqnarray}
  && 4\phi {\bar \psi}-c (\overline{f}^4-6\overline{f}^2 {\bar \psi}^2+{\bar \psi}^4) = 0,\label{eq:5c}\\
  && 4 \phi \overline{f}+4 c (\overline{f}^3 {\bar \psi} -\overline{f}{\bar \psi}^3) = 0,\label{eq:5d}
\end{eqnarray}
where the coefficient
$c = a +2/3$.

Because Eq.~(\ref{eq:5d}) is satisfied for arbitrarily small $\phi$, we immediately find $\alpha = \beta = 1/3$. However, to justify Eq.~(\ref{eq:25}) we must further prove that $k_{1,2}$ are real. To this end we substitute $\alpha = \beta = 1/3$ as well as Eq.~(\ref{eq:27}) into Eqs.~(\ref{eq:5c}) and (\ref{eq:5d}), obtaining
\begin{eqnarray}
  && 4 k_2-c (k_1^4-6 k_1^2 k_2^2+k_2^4) = 0,\label{eq:5e}\\
  && 4 k_1+4 c (k_1^3 k_2 - k_1 k_2) = 0.\label{eq:5f}
\end{eqnarray}
Upon eliminating the coefficient $c$ we reduce them to $k_1^4 -2 k_1^2 k_2^2 - 3 k_2^4 = 0$ which gives
\begin{equation}\label{eq:28}
    k_1^2 - 3 k_2^2 =0.
\end{equation}
Substituting it into Eq.~(\ref{eq:5f}) we obtain $k_1 =3^{1/2} (2 c)^{-1/3},k_2 = -(2c)^{-1/3}$. As a result,
\begin{eqnarray}
  && \overline{f} = 3^{1/2} (2 a + 4/3)^{-1/3} \phi^{1/3},\label{eq:5e}\\
  && {\bar \psi} = -(2 a + 4/3)^{-1/3} \phi^{1/3}.\label{eq:5f}
\end{eqnarray}
Summarizing, combined with numerical analysis we have shown that Eqs.~(\ref{eq:5e}) and (\ref{eq:5f}) are solutions to Eqs.~(\ref{eq:7}) and (\ref{eq:8}).

\section{Relation between field theory and universal eigenchannel structure}
\label{sec:diffuson}

In this Appendix we show that in the functional integral formalism, fluctuations around the saddle point carry information on the universal eigenchannel structure. To illustrate this we focus on the perfectly transmitting eigenchannel, i.e., the transmission eigenvalue ${\cal T}=1$ or equivalently $\phi=0$. Because of this the functional integral is reduced to
\begin{eqnarray}
\int_{(2z_b' Q\partial_x Q+[Q,\Lambda])|_{x=0}=0}^{(2z_b Q\partial_x Q - [Q,\Lambda])|_{x=L}=0} D[Q](\cdot) e^{-\frac{\xi}{8}
\int_0^L dx
{\rm str} (\partial_x Q)^2}.
\label{eq:58}
\end{eqnarray}
Here the pre-exponential factor $(\cdot)$ depends on specific observable considered. Its details are unimportant for present discussions of the general structure of field theory. Comparing with Eq.~(\ref{eq:3}), most importantly,
the boundary constraint of the right sample edge is replaced in Eq.~(\ref{eq:58}) by $(2z_b Q\partial_x Q - [Q,\Lambda])|_{x=L}=0$ (noting that $\Gamma|_{\theta=\phi=0}=\Lambda$) while that on the left remains the same.

To proceed further, we introduce the rational parametrization,
\begin{equation}\label{eq:59}
    Q=(1+iW)\Lambda (1+iW)^{-1},
\end{equation}
where the supermatrix $W\equiv\{W_{\alpha\alpha'}^{\lambda\lambda'}\}$ anticommutes with $\Lambda$, implying the ar-sector index $\lambda\neq \lambda'$. Recall that $\alpha,\alpha'$ are fb-sector indexes. Then, we expand $Q$ in terms of $W$ and substitute the expansion into both the action and the boundary constraints
(as well as the pre-exponential factor). For diffusive samples ($L\ll \xi$) we may keep the expansions up to the quadratic order. As a result, we reduce (\ref{eq:58}) to
\begin{eqnarray}
\int_{(z_b' \partial_x -1)W|_{x=0}=0}^{(z_b \partial_x +1)W|_{x=L}=0} D[W](\cdot) e^{-\frac{\xi}{2}
\int_0^L dx
{\rm str} (\partial_x W)^2},
\label{eq:60}
\end{eqnarray}
where we have used the fact that the Jacobian for the transformation: $Q\rightarrow W$ is unity \cite{Efetov97}.

The propagator of the effective field theory (\ref{eq:60}) is none but the well-known diffuson in diagrammatic techniques (recall that we do not consider time-reversal symmetry in the present work). More precisely, this propagator is given by \cite{Tian13a}
\begin{equation}\label{eq:61}
    \langle W_{\alpha\alpha'}^{\lambda\lambda'}(x) W_{\alpha'\alpha}^{\lambda'\lambda}(x')\rangle_0=\frac{1}{2\pi\nu}{\cal Y}_0(x,x').
\end{equation}
Here $\langle\cdot\rangle_0$ stands for the average with respect to the Gaussian weight (\ref{eq:60}) (subject to the boundary constraints), and $\nu$ is the average density of states. The localization length $\xi=2\pi\nu D_0$ with $D_0$ being the Boltzmann diffusion constant. Note that propagators with other index combinations vanish. Importantly, ${\cal Y}_0(x,x')$ is the solution to the normal diffusion equation:
\begin{eqnarray}
  -D_0\partial_x^2{\cal Y}_0(x,x') &=& \delta(x-x'), \label{eq:62}\\
  (z_b' \partial_x -1){\cal Y}_0(x,x')|_{x=0} &=& 0,\nonumber\\
  (z_b\partial_x+1){\cal Y}_0(x,x')|_{x=L} &=& 0, \nonumber
\end{eqnarray}
i.e.,
\begin{equation}\label{eq:63}
    {\cal Y}_0(x,x')=\frac{1}{D_0}\frac{(z_b'+x_<)(L+z_b-x_>)}{L+z_b+z_b'},
\end{equation}
where $x_{<(>)}={\rm min}({\rm max})\{x,x'\}$.

\begin{figure}[h]
  \centering
\includegraphics[width=8.0cm]{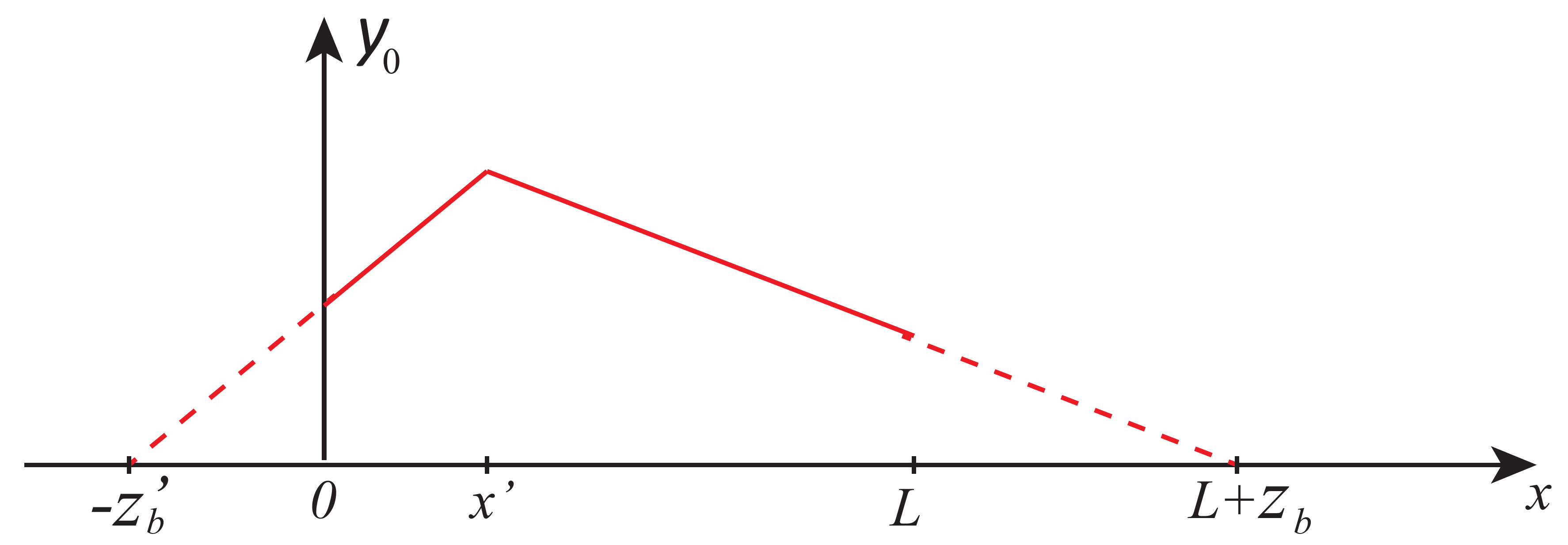}
 \caption{The energy density profile, ${\cal Y}_0(x,x')$, generated by injection of a unit flux at point $x'$ inside the sample.}
  \label{fig:10}
\end{figure}

Physically, ${\cal Y}_0(x,x')$ is the intensity (energy density) profile when a unit flux is injected at $x'$. The expression (\ref{eq:63}) entails the canonical meaning of $z_b,z_b'$ -- the so-called extrapolation length \cite{Rossum99}. As shown in Fig.~\ref{fig:10}, the profile linearly falls down from the source point and vanishes outside the sample and at a distance of $z_b'$ ($z_b$) to the left (right) end. In the limiting case of vanishing $z_b,z_b'$, as shown in Ref.~\onlinecite{Genack15}, ${\cal Y}_0(x,x'=x)$ is essentially $F_1(x)$ in Eq.~(\ref{eq:48}) up to an irrelevant overall factor and thereby determines the eigenchannel profile $W_{{\cal T}=1}(x)$. If we assume that the expression (\ref{eq:48}) is valid also for nonvanishing $z_b,z_b'$, then Eq.~(\ref{eq:49}) is received.

\section{Discussions on parametrization (\ref{eq:34})}
\label{sec:parametrization}

In this Appendix we show that provided $x$ is uniformly distributed in the range of $[0,L]$, Eq.~(\ref{eq:34}) is the unique parametrization leading to the bimodal distribution. For the convenience below we introduce the variable $\eta\equiv x/L$. We have
\begin{equation}\label{eq:39}
    \frac{d\eta}{d{\cal T}}=\frac{a'}{{\cal T}\sqrt{1-{\cal T}}},
\end{equation}
where the coefficient $a'$ is fixed by the normalization condition:
\begin{equation}\label{eq:40}
    \int_{T_{typ}}^1\frac{a'}{{\cal T}\sqrt{1-{\cal T}}}d{\cal T}=1.
\end{equation}
Since ${\cal T}(\eta)$ is symmetric with respect to the middle of the sample, $\eta$ is a double-valued function of ${\cal T}$. Solving Eq.~(\ref{eq:39}) gives
\begin{equation}\label{eq:41}
    \eta=\frac{a'}{2}\ln\left(\frac{1-\sqrt{1-{\cal T}}}{1+\sqrt{1-{\cal T}}}\frac{1+\sqrt{1-{\cal T}_{typ}}}{1-\sqrt{1-{\cal T}_{typ}}}\right).
\end{equation}
In combination with Eq.~(\ref{eq:40}) it gives
\begin{equation}\label{eq:42}
    2\left(\eta-\frac{1}{2}\right)\ln\frac{1+\sqrt{1-{\cal T}_{typ}}}{1-\sqrt{1-{\cal T}_{typ}}}=\pm\ln\frac{1-\sqrt{1-{\cal T}}}{1+\sqrt{1-{\cal T}}},
\end{equation}
where the $+$ ($-$) sign corresponds to $\eta<1/2$ ($>1/2$). From this expression we find
\begin{eqnarray}
  {\cal T} &=& \cosh^{-2} \left(b'\left(\eta-\frac{1}{2}\right)\right),
  \label{eq:43}\\
  b' &=& \ln\frac{1+\sqrt{1-{\cal T}_{typ}}}{1-\sqrt{1-{\cal T}_{typ}}}.
  \nonumber
\end{eqnarray}
Substituting $T_{typ} = e^{-2L/\ell}$ [cf. Eq.~(\ref{eq:38})] into the expression of $b'$ we find $b'\approx 2L/\ell$. Equation (\ref{eq:43}) then is identical to Eq.~(\ref{eq:34}).

\end{appendix}


\begin{thebibliography}{}
\bibitem{note1} Even in the absence of edge reflection Eq.~(\ref{eq:20}) has been shown \cite{Tian13} not to be exact. Instead, the sample length $L$ in Eq.~(\ref{eq:20}) must be replaced by $L+2z_{b0}$. However, provided that $z_{b0}\approx 0.7\ell$ is much smaller than $L$, this correction is negligible. Recall that $\ell$ is the transport mean free path.
\end{thebibliography}
\end{document}